\documentclass[prl,reprint,twocolumn,showpacs,superscriptaddress]{revtex4-1}
\usepackage{amsmath,amssymb,bm,mathrsfs,graphicx, braket, times}
\usepackage[colorlinks=true,citecolor=blue,linkcolor=blue]{hyperref}
\usepackage{longtable}
\usepackage{multirow}
\usepackage[all,cmtip]{xy}
\usepackage[normalem]{ulem}
\usepackage[usenames,dvipsnames]{color}

\renewcommand{\vec}[1]{\bm{#1}}

\newcommand{\sg}[1]{{\bf\em{#1}}}
\newcommand{\mS}{\mathcal{S}}
\newcommand{\mG}{\mathcal{G}}
\newcommand{\bN}{\mathbb{N}}

\begin{document}

\title{Filling-Enforced Gaplessness in Band Structures of the 230 Space Groups}

\author{Haruki Watanabe}
\affiliation{Department of Applied Physics, University of Tokyo, Tokyo 113-8656, Japan.}

\author{Hoi Chun Po}
\affiliation{Department of Physics, University of California, Berkeley, California 94720, USA.}

\author{Michael P. Zaletel}
\affiliation{Station Q, Microsoft Research, Santa Barbara, California, 93106, USA.}

\author{Ashvin Vishwanath}
\affiliation{Department of Physics, University of California, Berkeley, California 94720, USA.}
\affiliation{Materials Science Division, Lawrence Berkeley National Laboratories, Berkeley California 94720, USA.}

\begin{abstract}
Nonsymmorphic symmetries like screws and glides produce electron band touchings, obstructing the formation of a band insulator and leading, instead, to metals or nodal semimetals even when the number of electrons in the unit cell is an even integer. Here, we calculate the electron fillings compatible with being a band insulator for all 230 space groups, for noninteracting electrons with time-reversal symmetry. Our bounds are tight -- that is, we can rigorously eliminate band insulators at any forbidden filling and produce explicit models for all allowed fillings -- and stronger than those recently established for  interacting systems. These results provide simple criteria that should help guide the search for topological semimetals and, also, have implications for both the nature and stability of the resulting nodal Fermi surfaces. 
\end{abstract}

\maketitle

\paragraph{Introduction.}---
Recent advances in the understanding of topological phases of matter have rekindled the interest  in the interplay between electron filling and electronic phases of matter. When is a system of electrons insulating? 
For noninteracting electrons in the presence of time-reversal (TR) symmetry, basic band theory dictates that a band insulator (BI) is possible only if the electron filling $\nu$, defined as the average number of electrons per primitive unit cell, is an even integer. 

Since all crystals possess space group (SG) symmetries, it is of fundamental importance to ask whether the filling constraints are tightened due to the crystal structure, 
i.e.~do the extra spatial symmetries forbid BIs even when $\nu \in 2 \mathbb N$?
When spin-orbit coupling (SOC) is negligible, it has long been established that nonsymmorphic symmetries can enforce certain patterns of band degeneracies and lead to tighter filling constraints \cite{PhysRevB.56.13607,Zak2001}. 
Numerous recent works also pointed out that these nonsymmorphic filling constraints survive even when spin-rotation invariance is broken by SOC
\cite{
Young2012, PhysRevLett.112.036403, PhysRevB.92.081201, PhysRevB.91.205128, PhysRevB.92.165120, PhysRevLett.115.126803, 2015arXiv151208865C, 2015arXiv151201552F,PhysRevB.93.085427, 2015arXiv151200074W,NewFermions}. 
Weakly correlated materials forbidden to be insulating by such tightened filling constraints 
tend to favor nodal semimetals --- electronic systems with Fermi surfaces of reduced dimensionality consequentially feature low-energy excitations with unconventional dispersion. 

Similar filling constraints have also been derived for interacting systems using various nonperturbative methods 
\cite{Lieb1961, Oshikawa2000, Hastings2004, Sid2013, us1, Sid2015}.
However, none of the previous works provide tight constraints for all 230 SGs --- in our recent work on interacting systems \cite{us1}, we could only prove the tightness of the filling constraints for 218 SGs. That is, for the remaining 12 SGs at certain fillings, there was neither a general argument forbidding an insulator, nor an explicit construction of an insulating ground state.  

Here, we report the results from a comprehensive study of filling obstructions to realizing noninteracting TR-invariant BIs for all 230 SGs, with or without SOC.
Our key result is summarized in Table~\ref{summary}, which tabulates the set of electron fillings $\mS^{\text{BI}}_{\mG}$ compatible with a TR-symmetric BI in any given SG $\mG$. [See Tables~\ref{tabS1}, \ref{tabS2} of the Supplemental Materials (SM)~\footnote{See Supplemental Materials for expanded tables and discussions, which include Ref.~\cite{GroupTheory,Bradley,Bilbao,GK1,GK2}} for an expanded version]. Compared to the interacting results presented in \cite{us1}, the current Letter serves as an independent verification of the tight filling constraints for 218 SGs using band-theory analysis, and provides the tight bounds for the remaining 12 SGs in the noninteracting limit. In addition, the band-theory arguments presented here form the basis for further $\vec{k}\cdot \vec{p}$ effective Hamiltonian analysis, which constrains the generic dispersion about the degeneracy point \cite{PhysRevLett.108.266802, Yang2014, 2015arXiv150707504G, NewFermions}.
In contrast, our previous interacting argument
does not constrain the spectrum of low-energy excitation.

Before we move on to presenting the results, we comment on how such tight filling constraints for all possible crystal structures should be useful for materials design and screening. 
Materials with $\nu \not\in \mS_{\mG}^{\text{BI}}$ are necessarily (semi-)metallic or strongly interacting.  
At the same time, since the Luttinger volume of a system with any even filling $\nu=2n$ is zero, no Fermi surfaces are required~\cite{Luttinger,OshikawaLuttinger,Sid2015}. The simplest Fermiology that has a vanishing Luttinger volume, but, at the same time, is not an insulator, is a nodal point.   Such systems are attractive candidates for realizing nodal semimetals, although we should note that other outcomes involving compensated Fermi surfaces are also admissible. Therefore, the tight filling constraints we presented should be viewed as a general guide to help narrow down the search space to materials with a combination of SG symmetries and fillings that naturally favor nodal semimetals.

Actually, it may be worth noting that in spin-orbit-coupled systems lacking inversion symmetry, a similar argument on the Fermiology applies even when $\nu=2n+1$. 
Conventionally, this filling is associated with a large Fermi surface, encompassing half the Brillouin zone; in the presence of SOC, however, the individual spin components cannot be distinguished and the Luttinger volume constraint only applies to the total number of electrons. Hence, one could, in principle, realize a nodal semimetal at such fillings. However, if SOC is negligible or when the crystal is centrosymmetric, each band is doubly degenerate and the Luttinger's count is effectively halved.
Consequentially, unless both spatial inversion and spin-rotation symmetries are strongly broken, such systems typically possess two big Fermi surfaces each enclosing approximately half of the first Brillouin zone.

Let us mention some examples of existing materials that illustrate how our results apply. The proposed nodal-ring semimetal SrIrO$_{3}$ \cite{NComm_SrIrO3}, which has a topologically protected nodal Fermi surface \cite{PhysRevB.92.081201}, has $\nu  = 4$. For this SG ({\text{\sg{62}}}), however, the allowed BI fillings
are $ \mS^{\text{BI}}_{\text{\sg{62}}} = 8\mathbb N$, hence, the necessity of at least nodal points at the lower filling. 
Now consider the stability of this nodal structure to a symmetry-lowering distortion $\mG \rightarrow \mG'$. If $\mS^{\text{BI}}_{\mG} = \mS^{\text{BI}}_{\mG'}$, the nodal Fermi surface is guaranteed to be protected from a full gapping out, although it can change from, say, a collection of nodal lines to nodal points. Conversely, if $\nu \in  \mS^{\text{BI}}_{\mG'}$, a possibly nontrivial BI, such as a topological (crystalline) insulator, is in principle achievable via such distortion. 

\begin{table}
\caption{The list of fillings $\mS_{\mG}^{\text{BI}}$ corresponding to TR symmetric BIs in the presence or absence of SOC.
$m\bN$ represents the set $\{m, 2m, 3m, \cdots\}$ and $A \setminus B$ means deleting elements of $B$ from $A$.  $\mS_{\mG}^{\text{AI}}$'s are the corresponding fillings for 
AIs.  $\cap_{\Gamma \leq \mG} (\mS_{\Gamma}^{\text{BI}}/v_{\mG,\Gamma} )$ is the tightest constraints obtainable from Bieberbach subgroups $\Gamma$. \label{summary}}
\begin{tabular}{c|c|c|c}\hline\hline
$\mG$ (SG No.)						&$\mS_{\mG\times \rm{SU}(2)}^{\text{BI}} = \mS_{\mG}^{\text{AI}}$	& $\mS_{\mG}^{\text{BI}}$&$ \cap_{\Gamma \leq \mG} (\mS_{\Gamma}^{\text{BI}}/v_{\mG,\Gamma} )$	\\
\hline
\sg{1} & $2 \bN$  & $2 \bN$ & $2 \bN$ \\
\sg{4}, \sg{7}, \sg{9}   & $4 \bN$ & $4 \bN$ & $4 \bN$ \\
\sg{144}$\simeq$\sg{145} & $6 \bN$   & $6 \bN$ & $6 \bN$ \\
\sg{19}, \sg{29}, \sg{33}, \sg{76}$\simeq$\sg{78} & $8 \bN$  & $8 \bN$ & $8 \bN$ \\
\sg{169}$\simeq$\sg{170} & $12 \bN$ & $12 \bN$ & $12 \bN$ \\
\hline
\sg{73}, \sg{106}, \sg{110}, \sg{133},	&	\multirow{2}{*}{$8\bN$}			&	\multirow{2}{*}{$8\bN$} &	\multirow{2}{*}{$4\bN$} \\
\sg{135}, \sg{142}, \sg{206}, \sg{228}	& 			 			& 			&	 \\
\hline
\sg{199}, \sg{214}		&$4\bN\setminus\{4\}$				& $4\bN$								& $4\bN$	\\
\sg{220}					&$4\bN\setminus\{4,8,20\}$				& $4\bN\setminus\{4\}$					& $4\bN$	\\
\sg{230}					&$8\bN\setminus\{8\}$					& $8\bN$								& $4\bN$	\\
\hline
all other SGs& $2|\mathcal{W}^{\mG}_{a}|\bN$ 	&  $2|\mathcal{W}^{\mG}_{a}|\bN$ 	&  $2|\mathcal{W}^{\mG}_{a}|\bN$ \\
\hline\hline
\end{tabular}
\end{table}

\paragraph{Identification of special SGs.}---
We start by identifying some simple rules that reduce the analysis to a small number of special SGs.
Consider a class of system $X$ satisfying certain defining properties, like SG symmetries
. We will be interested in $\mS_{X}$, the set of electron fillings (defined with \emph{primitive} unit cell) for which a BI in class $X$ is possible. 
Imagine a ``less constrained" class $X'$ for which we lift some of the constraints imposed on $X$. By definition, a BI lying in class $X$ also lies in class $X'$, but the converse is not necessarily true. So the sets of fillings satisfy $\mS_{X}\subseteq\mS_{X'}$ provided the filling is defined with respect to the same unit cell on the two sides.

Such relations will greatly reduce the work required to establish the BI filling bounds for all 230 SGs. For instance, if SG $\mathcal G'$ is a subgroup of $\mathcal G$, systems symmetric under $\mG'$ belong to a less-constrained class compared to those symmetric under $\mG$. Therefore, we get an ``upper"-bound, $\mS^{\text{BI}}_{\mG}\subseteq  \mS^{\text{BI}}_{\mG'}/ v_{\mG, \mG'}$ for $\mG'<\mG$.  The factor $v_{\mG, \mG'} \geq 1$ is needed because $\mG'$ and $\mG$ may have different unit cell volumes.  For instance, if $\mG$ differs from $\mG'$ only by a body-centered translation, we have  $v_{\mG, \mG'} = 2$.  More generally, $v_{\mG, \mG'}= |T_{\mG}/ T_{\mG'}|$, where $T_{\mG}$ is the translation subgroup of $\mG$~\footnote{We write $H<G$ when $H$ is a proper subgroup of $G$, and $H\leq G$ when $H$ is possibly improper.  $|G/H|$ denotes the number of elements of the left coset $G/H$.}.

Atomic insulators (AIs) are special instances of BIs in which each electron is tightly localized to a single atomic orbital, or which can be smoothly deformed to such a configuration while preserving the symmetries.
This is a restriction on the phase, and thus, $\mS^{\text{AI}}_{\mathcal G}\subseteq\mS^{\text{BI}}_{\mathcal G}$ for the same $\mathcal G$, establishing a useful ``lower" bound.  Whenever the upper and lower bounds agree with each other, i.e., $\mS^{\text{AI}}_{\mathcal G} = \mS^{\text{BI}}_{\mathcal G'}/v_{\mG, \mG'}$ for some $\mathcal G' \leq \mathcal G$, one obtains the tight constraint $\mS^{\text{BI}}_{\mathcal G} = \mS^{\text{AI}}_{\mathcal G}$.

At first sight, it may appear nontrivial to deduce $\mS^{\text{BI}}_{\mathcal G}$ from this approach, since (i) one would still need to determine $\mS^{\text{AI}}_{\mathcal G}$, and (ii) there are numerous subgroup $\mathcal G'$ for any $\mathcal G$, and it is unclear which $\mathcal G'$ will provide maximal information about $\mS^{\text{BI}}_{\mathcal G}$. Fortunately (i) can be accomplished with little effort: by definition a TR-symmetric AI can be smoothly deformed into a phase with Kramers pairs of electrons localized to well-defined points in space, and these points must form a SG symmetric lattice. 
Such lattices are classified under the ``Wyckoff positions," which are exhaustively tabulated in \cite{ITC}.
Each Wyckoff position $\mathcal W^{\mG}_{w}$ ($w = a, b, \dots $) corresponds to a lattice with some number of points within each \emph{primitive} unit cell, which we denote by $|\mathcal W^{\mG}_{w}|$. Thus, $\mS^{\text{AI}}_{\mathcal G}$ is spanned by adding together arbitrary multiples of $2 |\mathcal W^{\mG}_{w}|$ (see Sec.~II of the SM).

For (ii), we take advantage of nonsymmorphic symmetries, which generally require extra band crossings and lead to tighter bounds on $\mS_{\mG}^{\text{BI}}$ \cite{Zak2001}.
To systematically study the effects of nonsymmorphic elements of $\mathcal{G}$, we first consider a special class of SGs that contains only screws, glides, and translations. 
When acting on $\mathbb{R}^3$, such groups are fixed-point free, i.e., $\vec{r}\neq g(\vec{r})$ for any pair of $\vec{r}\in\mathbb{R}^d$ and $g\in\mathcal{G}$ unless $g$ is the identity, and are known as ``Bieberbach" groups. Up to chirality, there are only ten such SGs in 3D \cite{ITC}: \sg{1}, \sg{4}, \sg{7}, \sg{9}, \sg{19}, \sg{29}, \sg{33}, \sg{76}($P4_1$)$\simeq$\sg{78}($P4_3$), \sg{144}($P3_1$)$\simeq$\sg{145}($P3_2$), \sg{169}($P6_1$)$\simeq$\sg{170}($P6_5$). 
(To avoid possible confusions, we denote a specific SG by its SG number used in \cite{ITC} in \sg{bold italic} face.)

In Sec.~IV of the SM, we establish that $\mS^{\text{AI}}_{\Gamma} = \mS^{\text{BI}}_{\Gamma}$ for each of the ten Bieberbach groups $\Gamma$ (the first five rows of Table~\ref{summary}). 
The ten Bieberbach groups then serve as an anchor for most of the analysis.
In particular, for many SGs $\mathcal G$, there exists a Bieberbach subgroup $\Gamma \leq \mathcal G$ satisfying $\mS^{\text{AI}}_{\mathcal G} = \mS^{\text{BI}}_{ \Gamma}/v_{\mG, \Gamma}$, which, therefore, establishes a tight bound on $\mS^{\text{BI}}_{\mG}$.
In fact, determining $\mS^{\text{AI}}_{\mG}$ and $\mS^{\text{BI}}_{\Gamma}$ allows us to derive $\mS_{\mG}^{\text{BI}}$ for 218 out of the 230 SGs [entries in Table~\ref{summary} with $\mS^{\text{AI}}_{\mG}=\mS^{\text{BI}}_{\mG} = \cap_{\Gamma \leq \mG} (\mS_{\Gamma}^{\text{BI}}/v_{\mG,\Gamma})$].
The 12 remaining SGs require a case-by-case study, though there are still group-subgroup relations among them that we can use to our advantage.
With that, our main result can be tersely summarized as follows. \emph{
For spinful electrons symmetric under both TR and SG $\mG$, $\mS_{\mG }^{\text{BI}} = \mS_{\mG }^{\text{AI}}$ unless $\mG$ belongs to one of the following four exceptions: \sg{199}, \sg{214}, \sg{220}, or \sg{230}.
}
These four exceptions allow for BIs even when no AI is possible at the same filling, and their topological properties are the focus of another study~\cite{us2}.  

Last, we address the case of systems with $\rm{SU}(2)$ spin-rotation invariance, relevant when SOC is negligible. The filling constraints satisfy
$\mS_{\mG\times \rm{SU}(2)}^{\text{AI}}\subseteq\mS_{\mG\times \rm{SU}(2)}^{\text{BI}}\subseteq\mS_{\mG}^{\text{BI}}$.  Since each entry in $\mS^{\text{AI}}_{\mG}$ corresponds to localizing an even number of electrons on each site, one can as well imagine putting them into a spin-singlet wave function, i.e., $\mS_{\mG\times \rm{SU}(2)}^{\text{AI}}  \,\, = \,\, \mS_{\mG}^{\text{AI}}$.
Again, by composing these relations, we have $\mS_{\mG\times\text{SU}(2)}^{\text{BI}} = \mS_{\mG }^{\text{BI}}$ whenever $\mS_{\mG }^{\text{BI}} = \mS_{\mG }^{\text{AI}}$. Hence, to establish the complete list of $\mS_{\mG\times\text{SU}(2)}^{\text{BI}} $ one simply studies the four SGs with $\mS^{\text{BI}}_{\mG}\neq \mS^{\text{AI}}_{\mG}$. The result is, \emph{for spinful electrons symmetric under TR, SG $\mG$, and spin rotation, $\mS_{\mG\times SU(2) }^{\text{BI}} = \mS_{\mG }^{\text{AI}}$ for all 230 SGs.}

\paragraph{Band theoretical analysis for \sg{73}.}---
We have reduced the analysis to that of the ten Bieberbach SGs $\Gamma$, and the 12 SGs for which knowledge on  $\mS^{\text{AI}}_{\mG}$ and $\mS^{\text{BI}}_{\Gamma}$ alone does not guarantee tightness of the bounds on $\mS_{\mG}^{\text{BI}}$. The filling constraints for some of these SGs, like those with a single screw or glide, have been derived in the literature \cite{
Zak2001, Young2012, PhysRevLett.112.036403, PhysRevB.92.081201, PhysRevB.91.205128, PhysRevB.92.165120, PhysRevLett.115.126803, 2015arXiv151208865C, 2015arXiv151201552F,PhysRevB.93.085427, 2015arXiv151200074W,NewFermions}. As briefly reviewed in Sec.~III of the SM, the main strategy is to derive the little group irreducible representations (irreps) at the high symmetry momenta, and, then, study the compatibility between irreps connected by high symmetry lines. Such an analysis, however, can become quite technical when the SG possesses a larger number of symmetries. Instead of presenting all the arguments for the $10+12$ SGs mentioned above, here, we focus only on \sg{73} in the presence of SOC, which illustrates the key ingredients needed to derive filling constraints for a SG with multiple nonsymmorphic symmetries and extra point group symmetries. We refer the interested readers to Secs.~IV and V of the SM for a detailed discussion on all the other SGs and the cases with spin-rotation invariance.

\sg{73} ($Ibca$) is centrosymmetric (i.e., contains the spatial inversion $P$) and belongs to the body-centered orthorhombic system.  
It is generated by $P$ and (two of the) three orthogonal screws 
$S_\alpha \equiv T_{\vec \tau_{\alpha} }R_{\alpha,\pi}$ with $\alpha = x,y,z$. $R_{\alpha,\theta}$ represents the anticlockwise rotation by angle $\theta$ around the positive $\alpha$ axis;  $T_{\vec{t}}$ represents the translation by $\vec{t}$, where, for the screws, we have $\vec \tau_{x} = (1/2,1/2,0)$, $\vec \tau_{y} = (0,1/2,1/2)$, and $\vec \tau_{z} = (1/2,0,1/2)$.

First, we study the constraints arising from the screws.
$S_z$ is a symmetry of the Bloch states along the line $\vec{k}=(\pi,\pi,k_z)$, which connects two high-symmetry points $\vec{k}=(\pi,\pi,0)$ (a time-reversal invariant momentum) and $(\pi,\pi,\pi)$ (\emph{not} time-reversal invariant due to the body-centered structure).  
Since $S_z^2=T_{(0,0,1)}R_{z,2\pi}$ and $2\pi$ rotation is $-1$ for a spin-$1/2$ electron, the allowed eigenvalues of $S_z$ along this line are $\xi^{(l)}_{z,  k_z} =  \xi^{(l)}_{z,0} e^{- i k_z/2} = \pm i e^{-ik_z/2}$ ($l$ is the band index). 
At $(\pi,\pi,0)$, bands with $\xi_{z,0}^{(l)} = \pm i$ are paired into Kramers doublets. Assuming a BI, the two bands forming a doublet at $(\pi,\pi,0)$ are either both filled or both empty. 
Therefore, along the line $(\pi,\pi,k_z)$ the number of filled bands having $\xi_{z,k_z}^{(l)} = i e^{- i k_z/2}$ will always be equal to that with $\xi_{z,k_z}^{(l)} = -i e^{- i k_z/2}$. 
The same argument, using the lines $(k_x,\pi,\pi)$ and $(\pi,k_y, \pi)$ applies equally well to $\xi_{x,k_x} $ and $\xi_{y,k_y}$.

\begin{figure}
\includegraphics[width=1\linewidth]{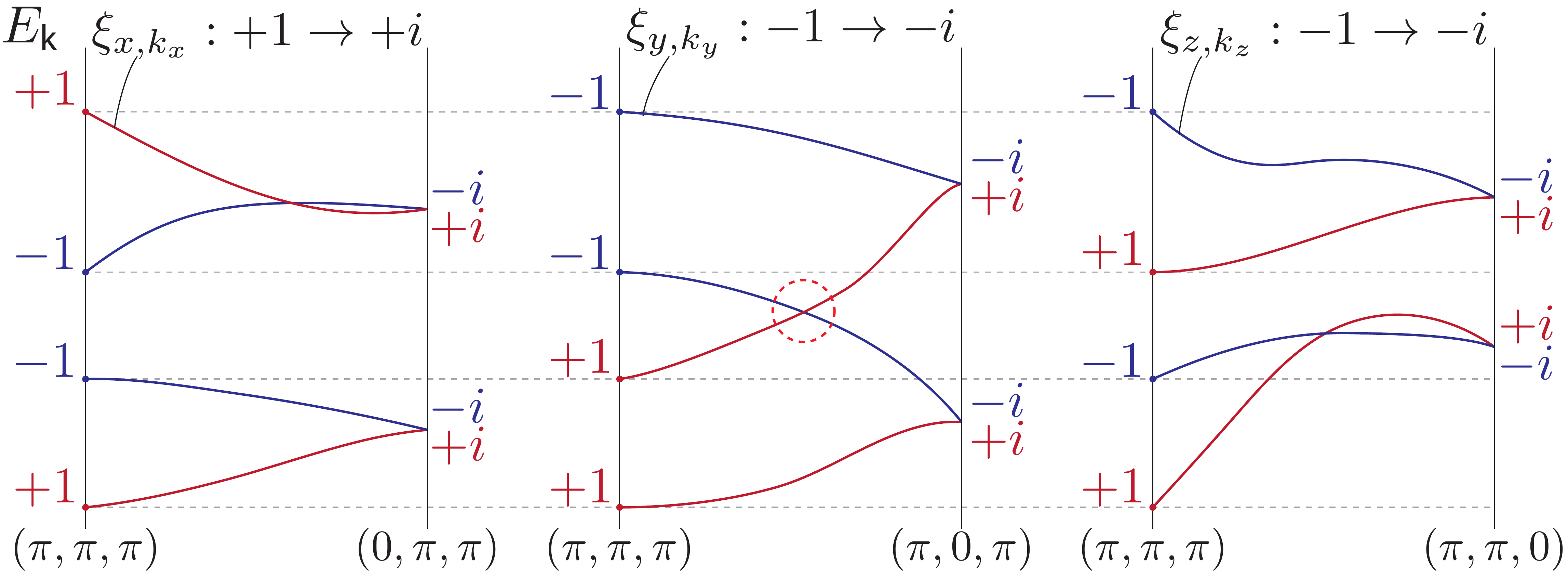}
\caption{Typical band structure of a TR-symmetric free electron system with SG \sg{73}. Each branch is doubly degenerate due to presence of TR and inversion. 
Note that the product $\xi_{x,\pi}^{(l)} \xi_{y,\pi}^{(l)} \xi_{z,\pi}^{(l)}$ is identical  for all bands (chosen to be $+1$ here).
The red dashed circle indicates inevitable crossings of four branches (a Dirac cone), enforcing $\nu= 8$. 
\label{fig:No73Fig}}
\end{figure}

Now, suppose there exists a BI at $\nu = 2$, and we focus on the symmetry representation at $(\pi,\pi,\pi)$ where all three screws are symmetries. 
On the one hand, the preceding discussion implies that the
two bands have opposite $\xi_{\alpha, \pi}$, i.e.~$\xi_{\alpha,\pi}^{(1)} \xi_{\alpha,\pi}^{(2)} = -1$ for each $\alpha = x,y,z$.
On the other hand, the screw eigenvalues of a single band are constrained by the group relations.
Since the product of three screws satisfies $S_x S_y S_z = 1$, we have $\xi_{x,\pi}^{(l)}\xi_{y,\pi}^{(l)}\xi_{z,\pi}^{(l)}=\pm1$. 
Note that the sign ambiguity, originating from the phase difference between $\pm \pi$ rotations on spin-1/2's, is independent of $l$.
Therefore, we require simultaneously
\begin{equation}\begin{split}
\left\{
\begin{array}{rl}
(\xi_{x,\pi}^{(1)}\xi_{x,\pi}^{(2)})(\xi_{y,\pi}^{(1)}\xi_{y,\pi}^{(2)})(\xi_{z,\pi}^{(1)}\xi_{z,\pi}^{(2)})&=(-1)^3 =-1\\
(\xi_{x,\pi}^{(1)} \xi_{y,\pi}^{(1)}\xi_{z,\pi}^{(1)})(\xi_{x,\pi}^{(2)} \xi_{y,\pi}^{(2)}\xi_{z,\pi}^{(2)})&=(\pm1)^2 =+1
\end{array}
\right.
,
\label{rep_pppz2}
\end{split}\end{equation}
a contradiction. More generally, the two conditions imply each of the four 1D irreps at $(\pi,\pi,\pi)$ appears the same number of times among the filled bands if the system is insulating, and therefore, $ \mS_{\text{\sg{73}}}^{\text{BI}} \subseteq 4 \mathbb N$ (Fig.~\ref{fig:No73Fig}).

To derive the tight bounds for $\mS_{\text{\sg{73}}}^{\text{BI}}$, however, one must utilize the inversion symmetry $P$, which was not assumed in the previous analysis.
As is well known, for spinful electrons, the combination of $P$ and $\mathcal T$ leads to doubly degenerate bands everywhere in the Brillouin zone. In particular $P \mathcal T$ commutes with $S_\alpha$ at $(\pi,\pi,\pi)$, and hence, the bands paired by $P \mathcal T$ have the same $\xi_{\alpha,\pi}^{(l)}$. The previous argument can then be applied to half of the bands (one from each pair). Combined with the observation that $ \mS_{\text{\sg{73}}}^{\text{AI}} = 8 \mathbb N$, we conclude $ \mS_{\text{\sg{73}}}^{\text{BI}} = 8 \mathbb N$.

\paragraph{Band insulators on flat manifolds.}---
Here we present an alternative derivation of $\mS_{\mG}^{\text{BI}}$ by defining the system on a nontrivial flat manifold. For simplicity we illustrate the main idea using $\mG=\text{\sg{73}}$ as an example, and a more general discussion is presented in Sec.~VI of the SM.

Suppose we are given a system of spinful electrons in $\mathbb{R}^3$ symmetric under $\mG=\text{\sg{73}}$.
Let us imagine putting the system on one of the ten compact flat manifolds in 3D. The most familiar example of such manifolds is the torus, which can be obtained by imposing periodic boundary conditions. In doing so, we (implicitly) take a translation subgroup $\Gamma^{(0)}$ of \sg{73} generated by $T_{\hat{\alpha}}^{L}=S_{\alpha}^{2L}$, and identify $\vec{r} \in \mathbb R^3$ by $\vec{r}\sim \vec{r} + L(l,m,n)$, with the two sides related by any $ T_{\hat x}^{L l}T_{\hat y}^{ L m}T_{\hat z}^{L n} = T_{L(l,m,n)}\in\Gamma^{(0)}$.  Here, $L$ is the linear dimension of the torus $\mathbb{R}^3/\Gamma^{(0)}$ and should be chosen much larger than the microscopic lattice constant $1$.  
To define the Hamiltonian on the torus, we also need to identify electronic creation operator $\hat{c}_i^{\dagger}(\vec{r})$ as  $\hat{c}^{\dagger}_i(\vec{r})\sim \hat{T}_{L(l,m,n)} \hat{c}^{\dagger}_i(\vec{r})\hat{T}_{L(l,m,n)}^{-1}=\hat{c}^{\dagger}_i( \vec{r} + L(l,m,n))$, where the subscript $i$ represents internal degrees of freedom.  

Replacing $\Gamma^{(0)}$ by other fixed-point-free subgroups of \sg{73} allows one to define the system on a nontrivial flat manifold.
Here, we choose a subgroup $\Gamma=\text{\sg{19}}$ generated by $\tilde{S}_\alpha \equiv T_{\vec \tau_{\alpha}}^L R_{\alpha,\pi}$ with an odd integer $L\gg1$. 
Note that, for instance, $\tilde{S}_x=T_{\vec \tau_{x}}^LR_{x,\pi}=(S_y)^{L-1}(S_x)^{L}$. 
The spatial points are identified as before, i.e.~$\vec r \sim \gamma(\vec r)$ for all $\gamma \in \Gamma$, and this gives the flat manifold $\mathcal{M}=\mathbb{R}^3/\Gamma$. The identification of operators are, however, nontrivial
\begin{equation}
\hat{c}^{\dagger}_i(\vec{r})\sim\hat{\gamma} \hat{c}^{\dagger}_i(\vec{r}) \hat{\gamma}^{-1} = \hat{c}^{\dagger}_j(\gamma(\vec{r})) (U_\gamma)_{ji}.\label{identification}
\end{equation}
Here, $U_{g}$ is a unitary representation of $g\in \mG$, and in contrast to the torus case, $U_{\gamma} \neq 1$, in general.  What is slightly complicated here is that the electron spin transforms projectively under spatial symmetries, i.e., $U_gU_{g'}=\omega_{g,g'}U_{gg'}$ for $g,g'\in\mG$, and $U_g$ intrinsically possesses sign ambiguity.  For example, $\pi$ rotation about the $\alpha$-axis can be represented by either of $ e^{\pm i\pi\frac{1}{2}\sigma_\alpha}=\pm i\sigma_\alpha$.  
To consistently identify operators, we need to fix the phase of $U_g$ in such a way that $U_{\gamma}$ ($\gamma \in \Gamma$) is a linear (nonprojective) representation of $\Gamma\subset\mG$. Such a choice of sign is always possible in 3D if $\Gamma$ is fixed-point free~\cite{Lutowski2014}. For the current problem, one can freely choose $U_{S_\alpha}= \pm i \sigma_\alpha$, but consistency demands $U_{T_{\alpha}}= U_{S_\alpha}^2 = -1$.

The argument presented so far uses only the Bieberbach subgroup $\Gamma$ and does not rely on the noninteracting assumption. Indeed, the interacting bounds we presented in \cite{us1} coincide with $\mS^{\text{BI}}_{\Gamma}/v_{\mG,\Gamma}$ for all SGs. As shown in Table~\ref{summary}, however, $\mS^{\text{BI}}_{\mG} \neq \mS^{\text{BI}}_{\Gamma}/v_{\mG,\Gamma} $ for ten SGs including \sg{73}. To derive the tight, noninteracting bounds for them, one must utilize the other SG symmetries differentiating $\mG$ from $\Gamma$.

Generally, an element $g$ in $\mG$ but not in $\Gamma$ may not remain a symmetry on $\mathcal{M}$. The necessary and sufficient condition for $g$ to remain a symmetry is that
\begin{eqnarray}
\forall\gamma\in\Gamma,\quad g\gamma g^{-1}\in\Gamma,\quad\text{and}\quad U_gU_\gamma U_g^{-1} = U_{g\gamma g^{-1}}.\label{remantsymm}
\end{eqnarray}
The first one is needed because $\vec{r}$ and $\gamma(\vec{r})$, the same point on $\mathcal{M}$, should be mapped to the same point again, i.e.~$g(\gamma(\vec{r}))\sim g(\vec{r})$. Similarly, the second one is to ensure that the operator identification is preserved: $ \hat g\hat{c}^{\dagger}_i(\vec{r}) \hat g^{-1} \sim \hat g\hat{\gamma}\hat{c}^{\dagger}_i(\vec{r}) \hat{\gamma}^{-1} \hat g^{-1}$.  For the problem at hand, both the rescaled body-centered translation $B=T_{\vec{\tau}}^L$ [$\vec{\tau}\equiv (1/2,1/2,1/2)$] and the  inversion $P$ are remnant symmetries.

To derive a filling constraint, we focus on the commutation relation of $B$ and $P$. They originally commute on $\mathbb{R}^3$, but here, we claim they must satisfy $\hat{B}\hat{P}=(-1)^{\hat{F}}\hat{P}\hat{B}$ as operators acting on Hilbert space, where $\hat{F}$ is the fermion number operator.  This follows from $B P = T_x T_y T_z P B$ and $U_{T_{\alpha}}=-1$.  Or, more explicitly, 
\begin{equation}\begin{split}
\hat{B}\hat{P}\hat{c}^{\dagger}_i(\vec{r})\hat{P}^{-1} \hat{B}^{-1}&= \hat{c}^{\dagger}_i(-\vec{r}+L\vec{\tau}); \\
\hat{P}\hat{B}\hat{c}^{\dagger}_i(\vec{r})\hat{B}^{-1}\hat{P}^{-1}&= \hat{c}^{\dagger}_i(-\vec{r}-L\vec{\tau}),
\end{split}\end{equation}
and the identification rule~\eqref{identification} implies $\hat{c}^{\dagger}_i(-\vec{r}+L\vec{\tau})\sim (-1)^3\hat{c}^{\dagger}_i(-\vec{r}-L\vec{\tau})$.

In addition to $B$ and $P$, the TR symmetry $\mathcal{T}$ [$\hat{\mathcal{T}}^2=(-1)^{\hat{F}}$] is also a remnant symmetry, and it still commutes with $B$ and $P$.  This algebra requires a four-fold degeneracy in the single particle spectrum, and hence we need $4n$ electrons on $\mathcal{M}$ to realize a BI. 
Since the number of unit cells contained in $\mathcal{M}$ is $\frac{L^3}{|\Gamma/T_{\Gamma}|}v_{\mG, \Gamma}=\frac{L^3}{2}$ (see Sec.~VI of the SM),  a BI is possible only if $\nu \frac{L^3}{2} \in 4 \mathbb N$. Recalling that $L$ is odd, one concludes $\mathcal{S}^{\text{BI}}_{\text{\sg{73}}}=8\mathbb{N}$.
Note that the nontriviality of the algebra hinges on $(-1)^{\hat F} = -1$ when acting on single-particle states, and therefore the obstruction can, in principle, be circumvented in the presence of interaction.

\paragraph{Outlook.}---
In this Letter, we have reported the full list of fillings for TR and SG symmetric BIs, with or without spin-rotation invariance.
The results also apply to 2D systems, since any layer group can be viewed as a ``slice" of a SG~\cite{ITC_E}. 
Understanding the nature of the enforced band degeneracies, the symmetry-topology protection of the nodal Fermi surfaces and the class of nontrivial BIs accessible by symmetry lowering are interesting open problems, as is the actual prediction of new materials candidates using these insights. 
As pointed out above, for 12 SGs, we could only prove tightness of the filling constraints in the noninteracting limit. It remains an interesting open problem whether interaction will enable trivial insulators at a lower filling in these systems.
Another promising future direction is the extension 
to magnetic SGs, pertinent for systems with magnetic ordering. 

\begin{acknowledgments}
The work of H. W. was mainly performed at Massachusetts Institute of Technology, and he acknowledges financial support from Pappalardo Fellowship.
H. C. P. is supported by a Hellman Graduate Award.
A. V.  acknowledges support from a Simons Investigator Grant. 
The work done at Berkeley (A. V. and H. C. P) is supported by NSF DMR-1411343.
\end{acknowledgments}

\bibliography{references}

\begin{thebibliography}{36}%
\makeatletter
\providecommand \@ifxundefined [1]{%
 \@ifx{#1\undefined}
}%
\providecommand \@ifnum [1]{%
 \ifnum #1\expandafter \@firstoftwo
 \else \expandafter \@secondoftwo
 \fi
}%
\providecommand \@ifx [1]{%
 \ifx #1\expandafter \@firstoftwo
 \else \expandafter \@secondoftwo
 \fi
}%
\providecommand \natexlab [1]{#1}%
\providecommand \enquote  [1]{``#1''}%
\providecommand \bibnamefont  [1]{#1}%
\providecommand \bibfnamefont [1]{#1}%
\providecommand \citenamefont [1]{#1}%
\providecommand \href@noop [0]{\@secondoftwo}%
\providecommand \href [0]{\begingroup \@sanitize@url \@href}%
\providecommand \@href[1]{\@@startlink{#1}\@@href}%
\providecommand \@@href[1]{\endgroup#1\@@endlink}%
\providecommand \@sanitize@url [0]{\catcode `\\12\catcode `\$12\catcode
  `\&12\catcode `\#12\catcode `\^12\catcode `\_12\catcode `\%12\relax}%
\providecommand \@@startlink[1]{}%
\providecommand \@@endlink[0]{}%
\providecommand \url  [0]{\begingroup\@sanitize@url \@url }%
\providecommand \@url [1]{\endgroup\@href {#1}{\urlprefix }}%
\providecommand \urlprefix  [0]{URL }%
\providecommand \Eprint [0]{\href }%
\providecommand \doibase [0]{http://dx.doi.org/}%
\providecommand \selectlanguage [0]{\@gobble}%
\providecommand \bibinfo  [0]{\@secondoftwo}%
\providecommand \bibfield  [0]{\@secondoftwo}%
\providecommand \translation [1]{[#1]}%
\providecommand \BibitemOpen [0]{}%
\providecommand \bibitemStop [0]{}%
\providecommand \bibitemNoStop [0]{.\EOS\space}%
\providecommand \EOS [0]{\spacefactor3000\relax}%
\providecommand \BibitemShut  [1]{\csname bibitem#1\endcsname}%
\let\auto@bib@innerbib\@empty
\bibitem [{\citenamefont {K\"onig}\ and\ \citenamefont
  {Mermin}(1997)}]{PhysRevB.56.13607}%
  \BibitemOpen
  \bibfield  {author} {\bibinfo {author} {\bibfnamefont {A.}~\bibnamefont
  {K\"onig}}\ and\ \bibinfo {author} {\bibfnamefont {N.~D.}\ \bibnamefont
  {Mermin}},\ }\href {\doibase 10.1103/PhysRevB.56.13607} {\bibfield  {journal}
  {\bibinfo  {journal} {Phys. Rev. B}\ }\textbf {\bibinfo {volume} {56}},\
  \bibinfo {pages} {13607} (\bibinfo {year} {1997})}\BibitemShut {NoStop}%
\bibitem [{\citenamefont {Michel}\ and\ \citenamefont {Zak}(2001)}]{Zak2001}%
  \BibitemOpen
  \bibfield  {author} {\bibinfo {author} {\bibfnamefont {L.}~\bibnamefont
  {Michel}}\ and\ \bibinfo {author} {\bibfnamefont {J.}~\bibnamefont {Zak}},\
  }\href {\doibase http://dx.doi.org/10.1016/S0370-1573(00)00093-4} {\bibfield
  {journal} {\bibinfo  {journal} {Phys. Rep.}\ }\textbf {\bibinfo {volume}
  {341}},\ \bibinfo {pages} {377 } (\bibinfo {year} {2001})}\BibitemShut
  {NoStop}%
\bibitem [{\citenamefont {Young}\ \emph {et~al.}(2012)\citenamefont {Young},
  \citenamefont {Zaheer}, \citenamefont {Teo}, \citenamefont {Kane},
  \citenamefont {Mele},\ and\ \citenamefont {Rappe}}]{Young2012}%
  \BibitemOpen
  \bibfield  {author} {\bibinfo {author} {\bibfnamefont {S.~M.}\ \bibnamefont
  {Young}}, \bibinfo {author} {\bibfnamefont {S.}~\bibnamefont {Zaheer}},
  \bibinfo {author} {\bibfnamefont {J.~C.~Y.}\ \bibnamefont {Teo}}, \bibinfo
  {author} {\bibfnamefont {C.~L.}\ \bibnamefont {Kane}}, \bibinfo {author}
  {\bibfnamefont {E.~J.}\ \bibnamefont {Mele}}, \ and\ \bibinfo {author}
  {\bibfnamefont {A.~M.}\ \bibnamefont {Rappe}},\ }\href {\doibase
  10.1103/PhysRevLett.108.140405} {\bibfield  {journal} {\bibinfo  {journal}
  {Phys. Rev. Lett.}\ }\textbf {\bibinfo {volume} {108}},\ \bibinfo {pages}
  {140405} (\bibinfo {year} {2012})}\BibitemShut {NoStop}%
\bibitem [{\citenamefont {Steinberg}\ \emph {et~al.}(2014)\citenamefont
  {Steinberg}, \citenamefont {Young}, \citenamefont {Zaheer}, \citenamefont
  {Kane}, \citenamefont {Mele},\ and\ \citenamefont
  {Rappe}}]{PhysRevLett.112.036403}%
  \BibitemOpen
  \bibfield  {author} {\bibinfo {author} {\bibfnamefont {J.~A.}\ \bibnamefont
  {Steinberg}}, \bibinfo {author} {\bibfnamefont {S.~M.}\ \bibnamefont
  {Young}}, \bibinfo {author} {\bibfnamefont {S.}~\bibnamefont {Zaheer}},
  \bibinfo {author} {\bibfnamefont {C.~L.}\ \bibnamefont {Kane}}, \bibinfo
  {author} {\bibfnamefont {E.~J.}\ \bibnamefont {Mele}}, \ and\ \bibinfo
  {author} {\bibfnamefont {A.~M.}\ \bibnamefont {Rappe}},\ }\href {\doibase
  10.1103/PhysRevLett.112.036403} {\bibfield  {journal} {\bibinfo  {journal}
  {Phys. Rev. Lett.}\ }\textbf {\bibinfo {volume} {112}},\ \bibinfo {pages}
  {036403} (\bibinfo {year} {2014})}\BibitemShut {NoStop}%
\bibitem [{\citenamefont {Fang}\ \emph {et~al.}(2015)\citenamefont {Fang},
  \citenamefont {Chen}, \citenamefont {Kee},\ and\ \citenamefont
  {Fu}}]{PhysRevB.92.081201}%
  \BibitemOpen
  \bibfield  {author} {\bibinfo {author} {\bibfnamefont {C.}~\bibnamefont
  {Fang}}, \bibinfo {author} {\bibfnamefont {Y.}~\bibnamefont {Chen}}, \bibinfo
  {author} {\bibfnamefont {H.-Y.}\ \bibnamefont {Kee}}, \ and\ \bibinfo
  {author} {\bibfnamefont {L.}~\bibnamefont {Fu}},\ }\href {\doibase
  10.1103/PhysRevB.92.081201} {\bibfield  {journal} {\bibinfo  {journal} {Phys.
  Rev. B}\ }\textbf {\bibinfo {volume} {92}},\ \bibinfo {pages} {081201}
  (\bibinfo {year} {2015})}\BibitemShut {NoStop}%
\bibitem [{\citenamefont {Gibson}\ \emph {et~al.}(2015)\citenamefont {Gibson},
  \citenamefont {Schoop}, \citenamefont {Muechler}, \citenamefont {Xie},
  \citenamefont {Hirschberger}, \citenamefont {Ong}, \citenamefont {Car},\ and\
  \citenamefont {Cava}}]{PhysRevB.91.205128}%
  \BibitemOpen
  \bibfield  {author} {\bibinfo {author} {\bibfnamefont {Q.~D.}\ \bibnamefont
  {Gibson}}, \bibinfo {author} {\bibfnamefont {L.~M.}\ \bibnamefont {Schoop}},
  \bibinfo {author} {\bibfnamefont {L.}~\bibnamefont {Muechler}}, \bibinfo
  {author} {\bibfnamefont {L.~S.}\ \bibnamefont {Xie}}, \bibinfo {author}
  {\bibfnamefont {M.}~\bibnamefont {Hirschberger}}, \bibinfo {author}
  {\bibfnamefont {N.~P.}\ \bibnamefont {Ong}}, \bibinfo {author} {\bibfnamefont
  {R.}~\bibnamefont {Car}}, \ and\ \bibinfo {author} {\bibfnamefont {R.~J.}\
  \bibnamefont {Cava}},\ }\href {\doibase 10.1103/PhysRevB.91.205128}
  {\bibfield  {journal} {\bibinfo  {journal} {Phys. Rev. B}\ }\textbf {\bibinfo
  {volume} {91}},\ \bibinfo {pages} {205128} (\bibinfo {year}
  {2015})}\BibitemShut {NoStop}%
\bibitem [{\citenamefont {Yang}\ \emph {et~al.}(2015)\citenamefont {Yang},
  \citenamefont {Morimoto},\ and\ \citenamefont
  {Furusaki}}]{PhysRevB.92.165120}%
  \BibitemOpen
  \bibfield  {author} {\bibinfo {author} {\bibfnamefont {B.-J.}\ \bibnamefont
  {Yang}}, \bibinfo {author} {\bibfnamefont {T.}~\bibnamefont {Morimoto}}, \
  and\ \bibinfo {author} {\bibfnamefont {A.}~\bibnamefont {Furusaki}},\ }\href
  {\doibase 10.1103/PhysRevB.92.165120} {\bibfield  {journal} {\bibinfo
  {journal} {Phys. Rev. B}\ }\textbf {\bibinfo {volume} {92}},\ \bibinfo
  {pages} {165120} (\bibinfo {year} {2015})}\BibitemShut {NoStop}%
\bibitem [{\citenamefont {Young}\ and\ \citenamefont
  {Kane}(2015)}]{PhysRevLett.115.126803}%
  \BibitemOpen
  \bibfield  {author} {\bibinfo {author} {\bibfnamefont {S.~M.}\ \bibnamefont
  {Young}}\ and\ \bibinfo {author} {\bibfnamefont {C.~L.}\ \bibnamefont
  {Kane}},\ }\href {\doibase 10.1103/PhysRevLett.115.126803} {\bibfield
  {journal} {\bibinfo  {journal} {Phys. Rev. Lett.}\ }\textbf {\bibinfo
  {volume} {115}},\ \bibinfo {pages} {126803} (\bibinfo {year}
  {2015})}\BibitemShut {NoStop}%
\bibitem [{\citenamefont {Chen}\ \emph {et~al.}(2016)\citenamefont {Chen},
  \citenamefont {Kim},\ and\ \citenamefont {Kee}}]{2015arXiv151208865C}%
  \BibitemOpen
  \bibfield  {author} {\bibinfo {author} {\bibfnamefont {Y.}~\bibnamefont
  {Chen}}, \bibinfo {author} {\bibfnamefont {H.-S.}\ \bibnamefont {Kim}}, \
  and\ \bibinfo {author} {\bibfnamefont {H.-Y.}\ \bibnamefont {Kee}},\ }\href
  {\doibase 10.1103/PhysRevB.93.155140} {\bibfield  {journal} {\bibinfo
  {journal} {Phys. Rev. B}\ }\textbf {\bibinfo {volume} {93}},\ \bibinfo
  {pages} {155140} (\bibinfo {year} {2016})}\BibitemShut {NoStop}%
\bibitem [{\citenamefont {Fang}\ \emph {et~al.}()\citenamefont {Fang},
  \citenamefont {Lu}, \citenamefont {Liu},\ and\ \citenamefont
  {Fu}}]{2015arXiv151201552F}%
  \BibitemOpen
  \bibfield  {author} {\bibinfo {author} {\bibfnamefont {C.}~\bibnamefont
  {Fang}}, \bibinfo {author} {\bibfnamefont {L.}~\bibnamefont {Lu}}, \bibinfo
  {author} {\bibfnamefont {J.}~\bibnamefont {Liu}}, \ and\ \bibinfo {author}
  {\bibfnamefont {L.}~\bibnamefont {Fu}},\ }\href
  {http://arxiv.org/abs/1512.01552} {\bibinfo  {journal} {arXiv:1512.01552}\
  }\BibitemShut {NoStop}%
\bibitem [{\citenamefont {Liang}\ \emph {et~al.}(2016)\citenamefont {Liang},
  \citenamefont {Zhou}, \citenamefont {Yu}, \citenamefont {Wang},\ and\
  \citenamefont {Weng}}]{PhysRevB.93.085427}%
  \BibitemOpen
\bibfield  {journal} {  }\bibfield  {author} {\bibinfo {author} {\bibfnamefont
  {Q.-F.}\ \bibnamefont {Liang}}, \bibinfo {author} {\bibfnamefont
  {J.}~\bibnamefont {Zhou}}, \bibinfo {author} {\bibfnamefont {R.}~\bibnamefont
  {Yu}}, \bibinfo {author} {\bibfnamefont {Z.}~\bibnamefont {Wang}}, \ and\
  \bibinfo {author} {\bibfnamefont {H.}~\bibnamefont {Weng}},\ }\href {\doibase
  10.1103/PhysRevB.93.085427} {\bibfield  {journal} {\bibinfo  {journal} {Phys.
  Rev. B}\ }\textbf {\bibinfo {volume} {93}},\ \bibinfo {pages} {085427}
  (\bibinfo {year} {2016})}\BibitemShut {NoStop}%
\bibitem [{\citenamefont {Wieder}\ \emph {et~al.}(2016)\citenamefont {Wieder},
  \citenamefont {Kim}, \citenamefont {Rappe},\ and\ \citenamefont
  {Kane}}]{2015arXiv151200074W}%
  \BibitemOpen
  \bibfield  {author} {\bibinfo {author} {\bibfnamefont {B.~J.}\ \bibnamefont
  {Wieder}}, \bibinfo {author} {\bibfnamefont {Y.}~\bibnamefont {Kim}},
  \bibinfo {author} {\bibfnamefont {A.~M.}\ \bibnamefont {Rappe}}, \ and\
  \bibinfo {author} {\bibfnamefont {C.~L.}\ \bibnamefont {Kane}},\ }\href
  {\doibase 10.1103/PhysRevLett.116.186402} {\bibfield  {journal} {\bibinfo
  {journal} {Phys. Rev. Lett.}\ }\textbf {\bibinfo {volume} {116}},\ \bibinfo
  {pages} {186402} (\bibinfo {year} {2016})}\BibitemShut {NoStop}%
\bibitem [{\citenamefont {Bradlyn}\ \emph {et~al.}(2016)\citenamefont
  {Bradlyn}, \citenamefont {Cano}, \citenamefont {Wang}, \citenamefont
  {Vergniory}, \citenamefont {Felser}, \citenamefont {Cava},\ and\
  \citenamefont {Bernevig}}]{NewFermions}%
  \BibitemOpen
  \bibfield  {author} {\bibinfo {author} {\bibfnamefont {B.}~\bibnamefont
  {Bradlyn}}, \bibinfo {author} {\bibfnamefont {J.}~\bibnamefont {Cano}},
  \bibinfo {author} {\bibfnamefont {Z.}~\bibnamefont {Wang}}, \bibinfo {author}
  {\bibfnamefont {M.}~\bibnamefont {Vergniory}}, \bibinfo {author}
  {\bibfnamefont {C.}~\bibnamefont {Felser}}, \bibinfo {author} {\bibfnamefont
  {R.~J.}\ \bibnamefont {Cava}}, \ and\ \bibinfo {author} {\bibfnamefont
  {B.~A.}\ \bibnamefont {Bernevig}},\ }\href
  {http://science.sciencemag.org/content/353/6299/aaf5037} {\bibfield
  {journal} {\bibinfo  {journal} {Science}\ }\textbf {\bibinfo {volume}
  {353}},\ \bibinfo {pages} {aaf5037} (\bibinfo {year} {2016})}\BibitemShut
  {NoStop}%
\bibitem [{\citenamefont {Lieb}\ \emph {et~al.}(1961)\citenamefont {Lieb},
  \citenamefont {Schultz},\ and\ \citenamefont {Mattis}}]{Lieb1961}%
  \BibitemOpen
  \bibfield  {author} {\bibinfo {author} {\bibfnamefont {E.}~\bibnamefont
  {Lieb}}, \bibinfo {author} {\bibfnamefont {T.}~\bibnamefont {Schultz}}, \
  and\ \bibinfo {author} {\bibfnamefont {D.}~\bibnamefont {Mattis}},\ }\href
  {\doibase 10.1016/0003-4916(61)90115-4} {\bibfield  {journal} {\bibinfo
  {journal} {Ann. Phys. (N.Y.)}\ }\textbf {\bibinfo {volume} {16}},\ \bibinfo
  {pages} {407 } (\bibinfo {year} {1961})}\BibitemShut {NoStop}%
\bibitem [{\citenamefont {Oshikawa}(2000{\natexlab{a}})}]{Oshikawa2000}%
  \BibitemOpen
  \bibfield  {author} {\bibinfo {author} {\bibfnamefont {M.}~\bibnamefont
  {Oshikawa}},\ }\href {\doibase 10.1103/PhysRevLett.84.1535} {\bibfield
  {journal} {\bibinfo  {journal} {Phys. Rev. Lett.}\ }\textbf {\bibinfo
  {volume} {84}},\ \bibinfo {pages} {1535} (\bibinfo {year}
  {2000}{\natexlab{a}})}\BibitemShut {NoStop}%
\bibitem [{\citenamefont {Hastings}(2004)}]{Hastings2004}%
  \BibitemOpen
  \bibfield  {author} {\bibinfo {author} {\bibfnamefont {M.~B.}\ \bibnamefont
  {Hastings}},\ }\href {\doibase 10.1103/PhysRevB.69.104431} {\bibfield
  {journal} {\bibinfo  {journal} {Phys. Rev. B}\ }\textbf {\bibinfo {volume}
  {69}},\ \bibinfo {pages} {104431} (\bibinfo {year} {2004})}\BibitemShut
  {NoStop}%
\bibitem [{\citenamefont {Parameswaran}\ \emph {et~al.}(2013)\citenamefont
  {Parameswaran}, \citenamefont {Turner}, \citenamefont {Arovas},\ and\
  \citenamefont {Vishwanath}}]{Sid2013}%
  \BibitemOpen
  \bibfield  {author} {\bibinfo {author} {\bibfnamefont {S.~A.}\ \bibnamefont
  {Parameswaran}}, \bibinfo {author} {\bibfnamefont {A.~M.}\ \bibnamefont
  {Turner}}, \bibinfo {author} {\bibfnamefont {D.~P.}\ \bibnamefont {Arovas}},
  \ and\ \bibinfo {author} {\bibfnamefont {A.}~\bibnamefont {Vishwanath}},\
  }\href {\doibase 10.1038/nphys2600} {\bibfield  {journal} {\bibinfo
  {journal} {Nat. Phys.}\ }\textbf {\bibinfo {volume} {9}},\ \bibinfo {pages}
  {299} (\bibinfo {year} {2013})}\BibitemShut {NoStop}%
\bibitem [{\citenamefont {Watanabe}\ \emph {et~al.}(2015)\citenamefont
  {Watanabe}, \citenamefont {Po}, \citenamefont {Vishwanath},\ and\
  \citenamefont {Zaletel}}]{us1}%
  \BibitemOpen
  \bibfield  {author} {\bibinfo {author} {\bibfnamefont {H.}~\bibnamefont
  {Watanabe}}, \bibinfo {author} {\bibfnamefont {H.~C.}\ \bibnamefont {Po}},
  \bibinfo {author} {\bibfnamefont {A.}~\bibnamefont {Vishwanath}}, \ and\
  \bibinfo {author} {\bibfnamefont {M.~P.}\ \bibnamefont {Zaletel}},\ }\href
  {http://www.pnas.org/content/112/47/14551.abstract} {\bibfield  {journal}
  {\bibinfo  {journal} {Proc. Natl. Acad. Sci. U.S.A.}\ }\textbf {\bibinfo
  {volume} {112}},\ \bibinfo {pages} {14551} (\bibinfo {year}
  {2015})}\BibitemShut {NoStop}%
\bibitem [{\citenamefont {Parameswaran}()}]{Sid2015}%
  \BibitemOpen
  \bibfield  {author} {\bibinfo {author} {\bibfnamefont {S.~A.}\ \bibnamefont
  {Parameswaran}},\ }\href {http://arxiv.org/abs/1508.01546} {\bibinfo
  {journal} {arXiv:1508.01546}\ }\BibitemShut {NoStop}%
\bibitem [{Note1()}]{Note1}%
  \BibitemOpen
\bibfield  {journal} {  }\bibinfo {note} {See Supplemental Materials for
  expanded tables and discussions, which include Ref.~\cite
  {GroupTheory,Bradley,Bilbao,GK1,GK2}}\BibitemShut {NoStop}%
\bibitem [{\citenamefont {Fang}\ \emph {et~al.}(2012)\citenamefont {Fang},
  \citenamefont {Gilbert}, \citenamefont {Dai},\ and\ \citenamefont
  {Bernevig}}]{PhysRevLett.108.266802}%
  \BibitemOpen
  \bibfield  {author} {\bibinfo {author} {\bibfnamefont {C.}~\bibnamefont
  {Fang}}, \bibinfo {author} {\bibfnamefont {M.~J.}\ \bibnamefont {Gilbert}},
  \bibinfo {author} {\bibfnamefont {X.}~\bibnamefont {Dai}}, \ and\ \bibinfo
  {author} {\bibfnamefont {B.~A.}\ \bibnamefont {Bernevig}},\ }\href {\doibase
  10.1103/PhysRevLett.108.266802} {\bibfield  {journal} {\bibinfo  {journal}
  {Phys. Rev. Lett.}\ }\textbf {\bibinfo {volume} {108}},\ \bibinfo {pages}
  {266802} (\bibinfo {year} {2012})}\BibitemShut {NoStop}%
\bibitem [{\citenamefont {Yang}\ and\ \citenamefont
  {Nagaosa}(2014)}]{Yang2014}%
  \BibitemOpen
  \bibfield  {author} {\bibinfo {author} {\bibfnamefont {B.-J.}\ \bibnamefont
  {Yang}}\ and\ \bibinfo {author} {\bibfnamefont {N.}~\bibnamefont {Nagaosa}},\
  }\href {http://dx.doi.org/10.1038/ncomms5898} {\bibfield  {journal} {\bibinfo
   {journal} {Nat. Commun.}\ }\textbf {\bibinfo {volume} {5}},\ \bibinfo
  {pages} {4898} (\bibinfo {year} {2014})}\BibitemShut {NoStop}%
\bibitem [{\citenamefont {Gao}\ \emph {et~al.}(2016)\citenamefont {Gao},
  \citenamefont {Hua}, \citenamefont {Zhang},\ and\ \citenamefont
  {Zhang}}]{2015arXiv150707504G}%
  \BibitemOpen
  \bibfield  {author} {\bibinfo {author} {\bibfnamefont {Z.}~\bibnamefont
  {Gao}}, \bibinfo {author} {\bibfnamefont {M.}~\bibnamefont {Hua}}, \bibinfo
  {author} {\bibfnamefont {H.}~\bibnamefont {Zhang}}, \ and\ \bibinfo {author}
  {\bibfnamefont {X.}~\bibnamefont {Zhang}},\ }\href {\doibase
  10.1103/PhysRevB.93.205109} {\bibfield  {journal} {\bibinfo  {journal} {Phys.
  Rev. B}\ }\textbf {\bibinfo {volume} {93}},\ \bibinfo {pages} {205109}
  (\bibinfo {year} {2016})}\BibitemShut {NoStop}%
\bibitem [{\citenamefont {Luttinger}(1960)}]{Luttinger}%
  \BibitemOpen
  \bibfield  {author} {\bibinfo {author} {\bibfnamefont {J.~M.}\ \bibnamefont
  {Luttinger}},\ }\href {\doibase 10.1103/PhysRev.119.1153} {\bibfield
  {journal} {\bibinfo  {journal} {Phys. Rev.}\ }\textbf {\bibinfo {volume}
  {119}},\ \bibinfo {pages} {1153} (\bibinfo {year} {1960})}\BibitemShut
  {NoStop}%
\bibitem [{\citenamefont {Oshikawa}(2000{\natexlab{b}})}]{OshikawaLuttinger}%
  \BibitemOpen
  \bibfield  {author} {\bibinfo {author} {\bibfnamefont {M.}~\bibnamefont
  {Oshikawa}},\ }\href {\doibase 10.1103/PhysRevLett.84.3370} {\bibfield
  {journal} {\bibinfo  {journal} {Phys. Rev. Lett.}\ }\textbf {\bibinfo
  {volume} {84}},\ \bibinfo {pages} {3370} (\bibinfo {year}
  {2000}{\natexlab{b}})}\BibitemShut {NoStop}%
\bibitem [{\citenamefont {Chen}\ \emph {et~al.}(2015)\citenamefont {Chen},
  \citenamefont {Lu},\ and\ \citenamefont {Kee}}]{NComm_SrIrO3}%
  \BibitemOpen
  \bibfield  {author} {\bibinfo {author} {\bibfnamefont {Y.}~\bibnamefont
  {Chen}}, \bibinfo {author} {\bibfnamefont {Y.-M.}\ \bibnamefont {Lu}}, \ and\
  \bibinfo {author} {\bibfnamefont {H.-Y.}\ \bibnamefont {Kee}},\ }\href
  {\doibase 10.1038/ncomms7593} {\bibfield  {journal} {\bibinfo  {journal}
  {Nat. Commun.}\ }\textbf {\bibinfo {volume} {6}},\ \bibinfo {pages} {6593}
  (\bibinfo {year} {2015})}\BibitemShut {NoStop}%
\bibitem [{Note2()}]{Note2}%
  \BibitemOpen
  \bibinfo {note} {We write $H<G$ when $H$ is a proper subgroup of $G$, and
  $H\leq G$ when $H$ is possibly improper. $|G/H|$ denotes the number of
  elements of the left coset $G/H$.}\BibitemShut {Stop}%
\bibitem [{\citenamefont {Hahn}(2006)}]{ITC}%
  \BibitemOpen
  \bibinfo {editor} {\bibfnamefont {T.}~\bibnamefont {Hahn}},\ ed.,\ \href@noop
  {} {\emph {\bibinfo {title} {International Tables for Crystallography}}},\
  \bibinfo {edition} {5th}\ ed.,\ Vol.\ \bibinfo {volume} {A: Space-group
  symmetry}\ (\bibinfo  {publisher} {Springer},\ \bibinfo {year}
  {2006})\BibitemShut {NoStop}%
\bibitem [{\citenamefont {Po}\ \emph {et~al.}(2016)\citenamefont {Po},
  \citenamefont {Watanabe}, \citenamefont {Zaletel},\ and\ \citenamefont
  {Vishwanath}}]{us2}%
  \BibitemOpen
  \bibfield  {author} {\bibinfo {author} {\bibfnamefont {H.~C.}\ \bibnamefont
  {Po}}, \bibinfo {author} {\bibfnamefont {H.}~\bibnamefont {Watanabe}},
  \bibinfo {author} {\bibfnamefont {M.~P.}\ \bibnamefont {Zaletel}}, \ and\
  \bibinfo {author} {\bibfnamefont {A.}~\bibnamefont {Vishwanath}},\ }\href
  {http://advances.sciencemag.org/content/2/4/e1501782} {\bibfield  {journal}
  {\bibinfo  {journal} {Sci. Adv.}\ }\textbf {\bibinfo {volume} {2}},\ \bibinfo
  {pages} {e1501782} (\bibinfo {year} {2016})}\BibitemShut {NoStop}%
\bibitem [{\citenamefont {Lutowski}\ and\ \citenamefont
  {Putrycz}()}]{Lutowski2014}%
  \BibitemOpen
  \bibfield  {author} {\bibinfo {author} {\bibfnamefont {R.}~\bibnamefont
  {Lutowski}}\ and\ \bibinfo {author} {\bibfnamefont {B.}~\bibnamefont
  {Putrycz}},\ }\href {http://arxiv.org/abs/1411.7799} {\bibinfo  {journal}
  {arXiv:1411.7799}\ }\BibitemShut {NoStop}%
\bibitem [{\citenamefont {Kopsk\'{y}}\ and\ \citenamefont
  {Litvin}(2010)}]{ITC_E}%
  \BibitemOpen
\bibfield  {journal} {  }\bibinfo {editor} {\bibfnamefont {V.}~\bibnamefont
  {Kopsk\'{y}}}\ and\ \bibinfo {editor} {\bibfnamefont {D.~B.}\ \bibnamefont
  {Litvin}},\ eds.,\ \href@noop {} {\emph {\bibinfo {title} {International
  Tables for Crystallography}}},\ \bibinfo {edition} {2nd}\ ed.,\ Vol.\
  \bibinfo {volume} {E: Subperiodic groups}\ (\bibinfo  {publisher}
  {Elsevier},\ \bibinfo {year} {2010})\BibitemShut {NoStop}%
\bibitem [{\citenamefont {Dresselhaus}\ \emph {et~al.}(2008)\citenamefont
  {Dresselhaus}, \citenamefont {Dresselhaus},\ and\ \citenamefont
  {Jorio}}]{GroupTheory}%
  \BibitemOpen
  \bibfield  {author} {\bibinfo {author} {\bibfnamefont {M.}~\bibnamefont
  {Dresselhaus}}, \bibinfo {author} {\bibfnamefont {G.}~\bibnamefont
  {Dresselhaus}}, \ and\ \bibinfo {author} {\bibfnamefont {A.}~\bibnamefont
  {Jorio}},\ }\href@noop {} {\emph {\bibinfo {title} {Group Theory}}}\
  (\bibinfo  {publisher} {Springer-Verlag, Berlin},\ \bibinfo {year}
  {2008})\BibitemShut {NoStop}%
\bibitem [{\citenamefont {Bradley}\ and\ \citenamefont
  {Cracknell}(1972)}]{Bradley}%
  \BibitemOpen
  \bibfield  {author} {\bibinfo {author} {\bibfnamefont {C.~J.}\ \bibnamefont
  {Bradley}}\ and\ \bibinfo {author} {\bibfnamefont {A.~P.}\ \bibnamefont
  {Cracknell}},\ }\href@noop {} {\emph {\bibinfo {title} {The Mathematical
  Theory of Symmetry in Solids: Representation Theory for Point Groups and
  Space Groups}}}\ (\bibinfo  {publisher} {Oxford University Press},\ \bibinfo
  {year} {1972})\BibitemShut {NoStop}%
\bibitem [{\citenamefont {Aroyo}\ \emph {et~al.}(2006)\citenamefont {Aroyo},
  \citenamefont {Kirov}, \citenamefont {Capillas}, \citenamefont {Perez-Mato},\
  and\ \citenamefont {Wondratschek}}]{Bilbao}%
  \BibitemOpen
  \bibfield  {author} {\bibinfo {author} {\bibfnamefont {M.~I.}\ \bibnamefont
  {Aroyo}}, \bibinfo {author} {\bibfnamefont {A.}~\bibnamefont {Kirov}},
  \bibinfo {author} {\bibfnamefont {C.}~\bibnamefont {Capillas}}, \bibinfo
  {author} {\bibfnamefont {J.~M.}\ \bibnamefont {Perez-Mato}}, \ and\ \bibinfo
  {author} {\bibfnamefont {H.}~\bibnamefont {Wondratschek}},\ }\href
  {http://www.cryst.ehu.es/rep/repres.html} {\bibfield  {journal} {\bibinfo
  {journal} {Acta Crystallogr. Sect. A}\ }\textbf {\bibinfo {volume} {62}},\
  \bibinfo {pages} {115} (\bibinfo {year} {2006})}\BibitemShut {NoStop}%
\bibitem [{\citenamefont {Karpilovsky}(1985)}]{GK1}%
  \BibitemOpen
  \bibfield  {author} {\bibinfo {author} {\bibfnamefont {G.}~\bibnamefont
  {Karpilovsky}},\ }\href@noop {} {\emph {\bibinfo {title} {Projective
  Representations of Finite Groups}}}\ (\bibinfo  {publisher} {Dekker, New
  York},\ \bibinfo {year} {1985})\BibitemShut {NoStop}%
\bibitem [{\citenamefont {Karpilovsky}(1993)}]{GK2}%
  \BibitemOpen
  \bibfield  {author} {\bibinfo {author} {\bibfnamefont {G.}~\bibnamefont
  {Karpilovsky}},\ }\href@noop {} {\emph {\bibinfo {title} {Group
  Representations}}},\ Vol.~\bibinfo {volume} {2}\ (\bibinfo  {publisher}
  {North-Holland, Amsterdam},\ \bibinfo {year} {1993})\BibitemShut {NoStop}%
\end{thebibliography}%

\clearpage

\onecolumngrid
\section{\large{Supplemental Materials}}
\vspace{1.0\baselineskip}
\twocolumngrid

\section{I.~~~Band insulator (BI) filling}
\label{SMI}
For reader's convenience, we reproduce Table~\ref{summary} of the main text in a more explicit form. Table~\ref{tab_app_bi1} is for the general case and \ref{tab_app_bi2} is for the SU(2) invariant case.

\begin{table}
\begin{center}
\caption{\label{tab_app_bi1} Fillings that realize a band insulator for each of 157 non-symmorphic space groups. Those space groups not listed here are symmorphic and hence $\nu=2n$. In this table, $n\in\bN$ can take any positive integral value.  $n^{+}$ for \sg{220} is an arbitrary integer greater than $1$.\label{tabS1}}
\begin{tabular}{|cc|cc|cc|cc|cc|cc|cc|}
No.	&$\nu$	&No.	&$\nu$	&No.	&$\nu$	&No.	&$\nu$	&No.	&$\nu$	&No.	&$\nu$	&No.	&$\nu$	\\\hline
\sg{4}&$4n$&\sg{39}&$4n$&\sg{66}&$4n$&\sg{100}&$4n$&\sg{129}&$4n$&\sg{165}&$4n$&\sg{201}&$4n$\\
\sg{7}&$4n$&\sg{40}&$4n$&\sg{67}&$4n$&\sg{101}&$4n$&\sg{130}&$8n$&\sg{167}&$4n$&\sg{203}&$4n$\\
\sg{9}&$4n$&\sg{41}&$4n$&\sg{68}&$4n$&\sg{102}&$4n$&\sg{131}&$4n$&\sg{169}&$12n$&\sg{205}&$8n$\\
\sg{11}&$4n$&\sg{43}&$4n$&\sg{70}&$4n$&\sg{103}&$4n$&\sg{132}&$4n$&\sg{170}&$12n$&\sg{206}&$8n$\\
\sg{13}&$4n$&\sg{45}&$4n$&\sg{72}&$4n$&\sg{104}&$4n$&\sg{133}&$8n$&\sg{171}&$6n$&\sg{208}&$4n$\\
\sg{14}&$4n$&\sg{46}&$4n$&\sg{73}&$8n$&\sg{105}&$4n$&\sg{134}&$4n$&\sg{172}&$6n$&\sg{210}&$4n$\\
\sg{15}&$4n$&\sg{48}&$4n$&\sg{74}&$4n$&\sg{106}&$8n$&\sg{135}&$8n$&\sg{173}&$4n$&\sg{212}&$8n$\\
\sg{17}&$4n$&\sg{49}&$4n$&\sg{76}&$8n$&\sg{108}&$4n$&\sg{136}&$4n$&\sg{176}&$4n$&\sg{213}&$8n$\\
\sg{18}&$4n$&\sg{50}&$4n$&\sg{77}&$4n$&\sg{109}&$4n$&\sg{137}&$4n$&\sg{178}&$12n$&\sg{214}&$4n$\\
\sg{19}&$8n$&\sg{51}&$4n$&\sg{78}&$8n$&\sg{110}&$8n$&\sg{138}&$8n$&\sg{179}&$12n$&\sg{218}&$4n$\\
\sg{20}&$4n$&\sg{52}&$8n$&\sg{80}&$4n$&\sg{112}&$4n$&\sg{140}&$4n$&\sg{180}&$6n$&\sg{219}&$4n$\\
\sg{24}&$4n$&\sg{53}&$4n$&\sg{84}&$4n$&\sg{113}&$4n$&\sg{141}&$4n$&\sg{181}&$6n$&\sg{220}&$4n^+$\\
\sg{26}&$4n$&\sg{54}&$8n$&\sg{85}&$4n$&\sg{114}&$4n$&\sg{142}&$8n$&\sg{182}&$4n$&\sg{222}&$4n$\\
\sg{27}&$4n$&\sg{55}&$4n$&\sg{86}&$4n$&\sg{116}&$4n$&\sg{144}&$6n$&\sg{184}&$4n$&\sg{223}&$4n$\\
\sg{28}&$4n$&\sg{56}&$8n$&\sg{88}&$4n$&\sg{117}&$4n$&\sg{145}&$6n$&\sg{185}&$4n$&\sg{224}&$4n$\\
\sg{29}&$8n$&\sg{57}&$8n$&\sg{90}&$4n$&\sg{118}&$4n$&\sg{151}&$6n$&\sg{186}&$4n$&\sg{226}&$4n$\\
\sg{30}&$4n$&\sg{58}&$4n$&\sg{91}&$8n$&\sg{120}&$4n$&\sg{152}&$6n$&\sg{188}&$4n$&\sg{227}&$4n$\\
\sg{31}&$4n$&\sg{59}&$4n$&\sg{92}&$8n$&\sg{122}&$4n$&\sg{153}&$6n$&\sg{190}&$4n$&\sg{228}&$8n$\\
\sg{32}&$4n$&\sg{60}&$8n$&\sg{93}&$4n$&\sg{124}&$4n$&\sg{154}&$6n$&\sg{192}&$4n$&\sg{230}&$8n$\\
\sg{33}&$8n$&\sg{61}&$8n$&\sg{94}&$4n$&\sg{125}&$4n$&\sg{158}&$4n$&\sg{193}&$4n$& &\\
\sg{34}&$4n$&\sg{62}&$8n$&\sg{95}&$8n$&\sg{126}&$4n$&\sg{159}&$4n$&\sg{194}&$4n$& &\\
\sg{36}&$4n$&\sg{63}&$4n$&\sg{96}&$8n$&\sg{127}&$4n$&\sg{161}&$4n$&\sg{198}&$8n$& &\\
\sg{37}&$4n$&\sg{64}&$4n$&\sg{98}&$4n$&\sg{128}&$4n$&\sg{163}&$4n$&\sg{199}&$4n$& &\\
\end{tabular}
\end{center}
\end{table}

\begin{table}
\begin{center}
\caption{\label{tab_app_bi2}Fillings that realize a SU(2)-invariant band insulator $\mS^{\text{BI}}_{\mG\times\text{SU}(2)}$ for each of 157 non-symmorphic space groups.  This table can also be understood as fillings for atomic insulators $\mS^{\text{AI}}_{\mG}$. 
$n^{+}\in\bN\setminus\{1\}$ and $n^{++}\in\bN\setminus\{1,2,5\}$. \label{tabS2}
}
\begin{tabular}{|cc|cc|cc|cc|cc|cc|cc|}
No.	&$\nu$	&No.	&$\nu$	&No.	&$\nu$	&No.	&$\nu$	&No.	&$\nu$	&No.	&$\nu$	&No.	&$\nu$	\\\hline
\sg{4}&$4n$&\sg{39}&$4n$&\sg{66}&$4n$&\sg{100}&$4n$&\sg{129}&$4n$&\sg{165}&$4n$&\sg{201}&$4n$\\
\sg{7}&$4n$&\sg{40}&$4n$&\sg{67}&$4n$&\sg{101}&$4n$&\sg{130}&$8n$&\sg{167}&$4n$&\sg{203}&$4n$\\
\sg{9}&$4n$&\sg{41}&$4n$&\sg{68}&$4n$&\sg{102}&$4n$&\sg{131}&$4n$&\sg{169}&$12n$&\sg{205}&$8n$\\
\sg{11}&$4n$&\sg{43}&$4n$&\sg{70}&$4n$&\sg{103}&$4n$&\sg{132}&$4n$&\sg{170}&$12n$&\sg{206}&$8n$\\
\sg{13}&$4n$&\sg{45}&$4n$&\sg{72}&$4n$&\sg{104}&$4n$&\sg{133}&$8n$&\sg{171}&$6n$&\sg{208}&$4n$\\
\sg{14}&$4n$&\sg{46}&$4n$&\sg{73}&$8n$&\sg{105}&$4n$&\sg{134}&$4n$&\sg{172}&$6n$&\sg{210}&$4n$\\
\sg{15}&$4n$&\sg{48}&$4n$&\sg{74}&$4n$&\sg{106}&$8n$&\sg{135}&$8n$&\sg{173}&$4n$&\sg{212}&$8n$\\
\sg{17}&$4n$&\sg{49}&$4n$&\sg{76}&$8n$&\sg{108}&$4n$&\sg{136}&$4n$&\sg{176}&$4n$&\sg{213}&$8n$\\
\sg{18}&$4n$&\sg{50}&$4n$&\sg{77}&$4n$&\sg{109}&$4n$&\sg{137}&$4n$&\sg{178}&$12n$&\sg{214}&$4n^{+}$\\
\sg{19}&$8n$&\sg{51}&$4n$&\sg{78}&$8n$&\sg{110}&$8n$&\sg{138}&$8n$&\sg{179}&$12n$&\sg{218}&$4n$\\
\sg{20}&$4n$&\sg{52}&$8n$&\sg{80}&$4n$&\sg{112}&$4n$&\sg{140}&$4n$&\sg{180}&$6n$&\sg{219}&$4n$\\
\sg{24}&$4n$&\sg{53}&$4n$&\sg{84}&$4n$&\sg{113}&$4n$&\sg{141}&$4n$&\sg{181}&$6n$&\sg{220}&$4n^{++}$\\
\sg{26}&$4n$&\sg{54}&$8n$&\sg{85}&$4n$&\sg{114}&$4n$&\sg{142}&$8n$&\sg{182}&$4n$&\sg{222}&$4n$\\
\sg{27}&$4n$&\sg{55}&$4n$&\sg{86}&$4n$&\sg{116}&$4n$&\sg{144}&$6n$&\sg{184}&$4n$&\sg{223}&$4n$\\
\sg{28}&$4n$&\sg{56}&$8n$&\sg{88}&$4n$&\sg{117}&$4n$&\sg{145}&$6n$&\sg{185}&$4n$&\sg{224}&$4n$\\
\sg{29}&$8n$&\sg{57}&$8n$&\sg{90}&$4n$&\sg{118}&$4n$&\sg{151}&$6n$&\sg{186}&$4n$&\sg{226}&$4n$\\
\sg{30}&$4n$&\sg{58}&$4n$&\sg{91}&$8n$&\sg{120}&$4n$&\sg{152}&$6n$&\sg{188}&$4n$&\sg{227}&$4n$\\
\sg{31}&$4n$&\sg{59}&$4n$&\sg{92}&$8n$&\sg{122}&$4n$&\sg{153}&$6n$&\sg{190}&$4n$&\sg{228}&$8n$\\
\sg{32}&$4n$&\sg{60}&$8n$&\sg{93}&$4n$&\sg{124}&$4n$&\sg{154}&$6n$&\sg{192}&$4n$&\sg{230}&$8n^{+}$\\
\sg{33}&$8n$&\sg{61}&$8n$&\sg{94}&$4n$&\sg{125}&$4n$&\sg{158}&$4n$&\sg{193}&$4n$& &\\
\sg{34}&$4n$&\sg{62}&$8n$&\sg{95}&$8n$&\sg{126}&$4n$&\sg{159}&$4n$&\sg{194}&$4n$& &\\
\sg{36}&$4n$&\sg{63}&$4n$&\sg{96}&$8n$&\sg{127}&$4n$&\sg{161}&$4n$&\sg{198}&$8n$& &\\
\sg{37}&$4n$&\sg{64}&$4n$&\sg{98}&$4n$&\sg{128}&$4n$&\sg{163}&$4n$&\sg{199}&$4n^{+}$& &\\
\end{tabular}
\end{center}
\end{table}

\section{II.~~~Atomic Insulator (AI) Filling}
\label{SMII}

Here we present the detail of the derivation of $\mS_{\mG}^{\text{AI}}$ for each of 230 space groups. 

Given a lattice compatible with the space group $\mG$, one can realize a symmetric TR and space-group symmetric AI by localizing a singlet pair of electrons per site. The filling $\nu$ of the AI is hence twice the number of lattice sites in the unit cell. 
As there may be several lattices for a single $\mG$, we have to classify distinct lattice types systematically. This is precisely what `Wyckoff positions' do.

In general, every lattice consistent with $\mG$ can be understood as a `crystallographic orbit' of an arbitrary point $\vec{r}_0$ in the lattice:
\begin{equation}
\Lambda_{\vec{r}_0}^{\mG}=\{g(\vec{r}_0)\,|\,g\in\mG\}.
\end{equation}
The structure of the lattice is specified by the `site symmetry group' of $\vec{r}_0$, which is the subgroup of $\mG$ that leaves $\vec{r}_0$ unmoved:
\begin{equation}
\mG_{\vec{r}_0}\equiv\{g\in\mG\,|\, g(\vec{r}_0)=\vec{r}_0\}.
\end{equation}
The larger $\mG_{\vec{r}_0}$ is, the fewer number of sites of $\Lambda_{\vec{r}_0}^{\mG}$ are there in a unit cell. 
The choice of $\vec{r}_0$ for a given lattice is not unique; if one choose $\vec{r}_0'=g(\vec{r}_0)$ ($g\in\mG$) instead, the site symmetry group will just be conjugated, i.e.~$\mG_{\vec{r}_0'}=g\mG_{\vec{r}_0}g^{-1}$.

A Wyckoff position groups several lattices in terms of the site symmetry group. Namely, two lattices $\Lambda_{\vec{r}_0}^{\mG}$ and $\Lambda_{\vec{r}_0'}^{\mG}$ are of the same type and belong to the same Wyckoff position if the site symmetry group of $\vec{r}_0$ and $\vec{r}_0'$ are conjugate with each other: $^\exists g\in\mG$ such that $\mG_{\vec{r}_0'}=g\mG_{\vec{r}_0}g^{-1}$. Therefore, in what follows, we use Wyckoff positions to refer to a specific lattice type.  One can find a full list of Wyckoff positions $\mathcal{W}^{\mG}_{w}$, labeled by the Wyckoff letter $w=a, b, c \ldots$, for each space group in Ref.~\cite{ITC}.  

Let $|\mathcal{W}^{\mG}_{w}|$ be the number of sites per unit cell of a lattice belonging to the Wyckoff position $\mathcal{W}^{\mG}_{w}$.  Wyckoff positions are ordered in such a way that $|\mathcal{W}^{\mG}_{a}|\leq|\mathcal{W}^{\mG}_{b}|\leq|\mathcal{W}^{\mG}_{c}|\leq\cdots$.  Among the 230 space groups, 73 are symmorphic and $|\mathcal{W}^{\mG}_{a}|=1$, while 157 are nonsymmorphic and $|\mathcal{W}^{\mG}_{a}|\geq2$. 

As explained above, one can realize a symmetric AI at filling $\nu=2|\mathcal{W}^{\mG}_{w}|$.
Since superpositions of these AIs with arbitrary positive integer coefficients are again AIs, in general we have
\begin{equation}
\mS_{\mG}^{\text{AI}}=\{\sum_{w}2|\mathcal{W}^{\mG}_{w}|\,n_w\,\,|\,\,n_{w}\in\mathbb{N}\}.
\label{app:generalAI}
\end{equation}
For almost all of the 230 space groups, every $|\mathcal{W}^{\mG}_{w}|$ ($w\neq a$) is an integer multiple of $|\mathcal{W}^{\mG}_{a}|$, so 
\begin{equation}
\mS_{\mG}^{\text{AI}}=2m\bN \equiv\{2m, 4m, 6m, \cdots\} \,\,\text{where}\,\, m=|\mathcal{W}^{\mG}_{a}|.
\end{equation}
The exceptions are the four `Wyckoff-mismatched' space groups (\sg{199}, \sg{214}, \sg{220}, and \sg{230}).  For example, \sg{199} has three Wyckoff positions: $|\mathcal{W}^{\text{\sg{199}}}_{a}|=4$, $|\mathcal{W}^{\text{\sg{199}}}_{b}|=6$, and $|\mathcal{W}^{\text{\sg{199}}}_{c}|=12$.  Therefore, from Eq.~\eqref{app:generalAI}, spinless electrons can realize an AI at filling $\nu=8n_{a}+12n_{b}+24n_{c}$ ($n_{a},n_{b},n_{c}\in \mathbb{N}$), which altogether form a set $4\bN\setminus\{4\}\equiv \{8,12,16,20\ldots\}$.  Other three space groups can be discussed in the same way.

The atomic insulator fillings, derived in this way, turn out to be identical to those for SU(2)-invariant band insulators in Table~\ref{tab_app_bi2}.

\section{III.~~~Representations of Bloch states}
\label{SMIII}

\subsection{A.~~~Little Group Representations}
As preparation for the band theoretical analysis in the next section, here we review the little group representation of a space group $\mG$.  

Let us write a symmetry element $g\in\mathcal{G}$ as $g=\{p_g|\vec{t}_g\}$ when $g(\vec{r})=p_g\vec{r}+\vec{t}_g$. In general, $g=\{p_g|\vec{t}_g\}$ changes the lattice momentum $\vec{k}$ to $p_g\vec{k}$.  The subgroup of $\mathcal{G}$ that preserves $\vec{k}$ up to reciprocal lattice vectors $\vec G$ is called the little group of $\vec{k}$ (or the $\vec{k}$-group): 
\begin{equation}
\mathcal{G}_{\vec{k}}\equiv\{g\in \mathcal{G}|p_g\vec{k}=\vec{k}+{^\exists}\vec{G}\}.
\end{equation}
Since the single particle Hamiltonian $H_{\vec{k}}$ is invariant under $\mathcal{G}_{\vec{k}}$, Bloch states in general form a representation of $\mathcal{G}_{\vec{k}}$~\cite{GroupTheory}. By decomposing the representation into `irreps' (irreducible representations), one can obtain crucial information about the band structure such as the degeneracy at $\vec{k}$. Therefore, it is important to know all the irreps of $\mathcal{G}_{\vec{k}}$.  It is however troublesome to examine $\mathcal{G}_{\vec{k}}$ itself since $\mathcal{G}_{\vec{k}}$ is an infinite group because of its lattice translation subgroup $T_{\mathcal{G}}\sim\mathbb{Z}^d$.  Hence, as a step to derive irreps of $\mathcal{G}_{\vec{k}}$, one can discuss a finite group $P_{\vec{k}}\equiv\mathcal{G}_{\vec{k}}/T_{\mathcal{G}}$ instead.

There are two origins that make relevant representations of $P_{\vec{k}}$ generically projective.
Let $u_{\vec{k}}(p)$ be the representation of $p \in P_{\vec k}$, then
$u_{\vec{k}}(p)u_{\vec{k}}(p')=\omega_{\vec{k}}(p,p')u_{\vec{k}}(pp')$ for $p,p'\in P_{\vec{k}}$. See Sec.~VII for the review of projective representations.  

The first origin is the electron spin.  
When acting on spatial coordinates, we express $p$ as $p=e^{-i\vec{L}\cdot\vec{\theta}_{p}}$ for proper rotations and $p=-e^{-i\vec{L}\cdot\vec{\theta}_{p}}$ for improper rotations using the three-dimensional matrix representation of the angular momentum $\vec{L}$. ($\vec{\theta}_{p}$ is of course ambiguous but we arbitrarily choose and fix it for each $p$.)  
$\vec L$ above is replaced by $\vec \sigma$ (the vector of Pauli matrices) when $p$ acts on electron spin.
The factor system arising from the spin can be computed by
\begin{eqnarray}
z(p,p')\equiv e^{-\frac{i}{2}\vec{\sigma}\cdot\vec{\theta}_{p}}e^{-\frac{i}{2}\vec{\sigma}\cdot\vec{\theta}_{p'}}e^{\frac{i}{2}\vec{\sigma}\cdot\vec{\theta}_{pp'}}=\pm1.\label{app:pr_spin}
\end{eqnarray}

The second origin is the non-symmorphic nature of $\mathcal{G}_{\vec{k}}$. When the space group is symmorphic, one can always set $\vec{t}_g=0$ modulo lattice translations for all $g\in \mathcal{G}$ by properly choosing the origin. This is not the case for non-symmorphic space groups and nonzero $\vec{t}_g$ results in a factor system
\begin{equation}
\rho_{\vec{k}}(p_g,p_{g'})\equiv e^{i\vec{k}\cdot(\vec{t}_{g'}-p_g\vec{t}_{g'})}= e^{i(\vec{k}-p_g^{-1}\vec{k})\cdot\vec{t}_{g'}}.\label{app:factor2}
\end{equation}
As we are discussing $P_{\vec{k}}=\mathcal{G}_{\vec{k}}/T_{\mathcal{G}}$ rather than $\mathcal{G}_{\vec{k}}$, $\vec{t}_{g'}$ in Eq.~\eqref{app:factor2} is defined only up to a primitive lattice vector. However, this ambiguity does not affect $\rho(p_g,p_{g'})$ since $\vec{k}-p_g^{-1}\vec{k}$ is a reciprocal lattice vector $\vec{G}$. If $\vec{k}$ is inside of the Brillouin zone, $\vec{k}-p_g^{-1}\vec{k}=0$ and $\rho_{\vec{k}}=1$.

The factor system $\omega_{\vec{k}}$ above is the product of $z$ and $\rho_{\vec{k}}$,
\begin{equation}
\omega_{\vec{k}}(p,p')\equiv z(p,p')\rho_{\vec{k}}(p,p').
\end{equation}
Once we get irreps $u_{\vec{k}}^{(a)}$ of $P_{\vec{k}}$, we immediately know all irreps $U_{\vec{k}}^{(a)}(g)\equiv u_{\vec{k}}^{(a)}(p_g)e^{-i\vec{k}\cdot\vec{t}_g}$ of $\mG_{\vec{k}}$, which satisfies
\begin{eqnarray}
U_{\vec{k}}^{(a)}(g)U_{\vec{k}}^{(a)}(g')=z(p_g,p_g')U^{(a)}_{\vec{k}}(gg'),\,\, g,g'\in \mG_{\vec{k}}.
\end{eqnarray}

\subsection{B.~~~Time-Reversal Symmetry}
\label{app:TR}
In addition to the space group $\mG$, we assume that the system has time-reversal symmetry $\mathcal{T}$. In general, $\mathcal{T}$ is represented by a unitary matrix $U_{\mathcal{T}}$ followed by the complex conjugation $K$. $U_{\mathcal{T}}$ is antisymmetric when $\mathcal{T}^2=-1$.  Furthermore, since $g\in\mG$ and $\mathcal{T}$ commute, we need 
\begin{equation}
U_{\mathcal{T}}U_{\vec{k}}^*(g)U_{\mathcal{T}}^{-1}=U_{\vec{k}}(g)\label{app:UTR}
\end{equation}
for all $g\in\mG$ a TRIMs.

\subsection{C.~~~Spin SU(2) rotation}
\label{app:SOC}
A space group element $g\in\mG$ permutes the coordinate $\vec{r}$ of electrons and at the same time rotates the spin of electrons.
In the presence of spin-rotation symmetry, we can `undo' the spin rotation part of $g$ by using the corresponding element of SU(2).
Therefore the relevant band theory effectively reduces to the one for spinless electrons.  Consequently, one should set $z(p,p')=1$ in the above equations.  Irreps $u_{\vec{k}}^{(a)}$ of $P_{\vec{k}}$ may still be projective due to nontrivial $\rho_{\vec{k}}(p,p')$ when $\mG$ is nonsymmorphic.

\section{IV.~~~Band theory derivation of $S_{\mG}^{\text{BI}}$ for 10 fixed-point-free space groups $\Gamma$}
\label{SMIV}
Here we present band theoretical derivation of $S_{\mG}^{\text{BI}}$ for the 10 fixed-point-free (i.e.~Bieberbach) space groups in Table~\ref{tab_app_uc}. 

\begin{table}
\begin{center}
\caption{\label{tab_app_uc} Some properties of 10 fixed-point-free space groups $\Gamma$. $T_{\Gamma}$ is the translation subgroup of
$\Gamma$ and $|\Gamma/T_{\Gamma}|$ is the number of elements of the quotient group $\Gamma/T_{\Gamma}$.}
\begin{tabular}{c|c|c|c}
No. &key elements& $|\Gamma/T_{\Gamma}|$& $S_{\mG}^{\text{BI}}=S_{\mG}^{\text{AI}}$\\\hline
\sg{1} & -- & $1$ & $2\mathbb{N}$\\
\sg{4} &$2_1$ & $2$& $4\mathbb{N}$ \\ 
\sg{7} & glide& $2$ & $4\mathbb{N}$\\
\sg{9} & glide & $2$ & $4\mathbb{N}$\\
\sg{19} &$2_1 2_1$& $4$ & $8\mathbb{N}$\\
\sg{29} &glide+$2_1$& $4$ & $8\mathbb{N}$\\
\sg{33} &glide+$2_1$& $4$ & $8\mathbb{N}$\\
\sg{76}, \sg{78} 	&  $4_1$, $4_3$ & $4$ & $8\mathbb{N}$\\
\sg{144}, \sg{145} &$3_1$, $3_2$& $3$ & $6\mathbb{N}$\\ 
\sg{169}, \sg{170} &$6_1$, $6_5$& $6$ & $12\mathbb{N}$
\end{tabular}
\end{center}
\end{table}

\subsection{A.~~~Bieberbach space groups with a single screw or glide}
Let us start with those space groups which have only a $n_m$-screw axis or a glide in addition to lattice translations. This class includes 7 of the 10 fixed-point-free groups. The remaining three, \sg{19}, \sg{29}, and \sg{33}, are discussed later separately.

To discuss these space groups at once, we choose the primitive lattice vector $\vec{a}_{1,2,3}$ in such a way that (i) the screw or the glide $g$ is represented as 
\begin{equation}
g=e^{-i(m/n)\vec{P}\cdot\vec{a}_1} X.
\end{equation}
Here, $X$ is the rotation about an axis parallel to $\vec{a}_1$ for $n_m$-screws, or the mirror about a plane that contains $\vec{a}_1$ for a glide (where $m/n=1/2$).
(ii) The plane spanned by $\vec{a}_2$ and $\vec{a}_3$ is invariant under $\hat{X}$.
For these elementary groups, one can check case by case that such a choice is possible.  $g$ is then a symmetry of the single-particle Hamiltonian $h_{\vec{k}}$ along the line
\begin{equation}
\vec{k}=\kappa \vec{b}_1,\quad \kappa\in[-0.5,0,5],
\end{equation}
and so one can simultaneously diagonalize $h_{\vec{k}}$ and $g$. 

For example, in the case of $2_1$-screw for \sg{4}, $g^2$ is $2\pi$-rotation followed by translation $e^{-i\vec{P}\cdot\vec{a}_1}$. Thus, $g$ must be represented by
\begin{equation}
U_{\vec{k}}(g)^2=(-1)\times e^{-i\vec{k}\cdot\vec{a}_1}=-e^{-2\pi i\kappa}
\end{equation}
for spinful electrons.  There are two 1D reps along this line:
\begin{equation}
U_{\vec{k}}^{(\pm)}(g)=\pm i e^{-\pi i\kappa}.
\end{equation}
If one tracks $U_{\vec{k}}^{(\pm)}$ as $\kappa$ increases from $-0.5$ to $+0.5$, one finds that they acquire a negative sign and the two representations interchange with each other: $U_{\kappa=-0.5}^{(\pm)}=U_{\kappa=+0.5}^{(\mp)}$. However, $\kappa=-0.5$ and $+0.5$ are the same point in BZ.  The only resolution is that the same number of $U_{\vec{k}}^{(+)}$ and $U_{\vec{k}}^{(-)}$ appear and intersect with each other somewhere along this line.

Furthermore, since time-reversal $\mathcal{T}$ and $g$ commute, a Kramers pair at a TRIM must have eigenvalues of $g$ that are complex conjugate to each other.  At $\kappa=0.5$, however, $U_{\vec{k}}^{(+)}(g)=\pm1$ is real and thus there must be $2n$ copies of $U_{\vec{k}}^{(+)}$ to form Kramers pairs. Since there must be 
an equal number of bands corresponding to $U_{\vec{k}}^{(-)}$, one can conclude any set of isolated bands must contain $4n$ bands, i.e., $S_{\text{\sg{4}}}^{\text{BI}}\subseteq 4\bN$.  The tightness of the bound ($\mS^{\text{BI}}_{\text{\sg{4}}} = 4\bN$) can be shown by comparing with $\mS^{\text{AI}}_{\text{\sg{4}}}=4\bN$.

Other space groups with a single screw can be argued in a similar manner.
The same argument can also be applied to \sg{7} and \sg{9}, which contain a single glide (a mirror followed by a half translation). 
This is because the mirror operation by a plane normal to $\vec{n}$ is defined as $\pi$-rotation around $\vec{n}$ followed by the 3D inversion. Since the inversion does not transform spin at all, the discussion is unchanged from the $2_1$-screw case.

\subsection{B.~~~Bieberbach space groups with more than one nonsymmorphic elements}
Now we discuss the slightly more nontrivial examples for which the optimal bound can only be found by considering more than one symmetry operations.  
\subsubsection{\sg{19}}
The condition for \sg{19} can actually be deduced from our discussion on \sg{73} presented in the main text.  \sg{19} belongs to the primitive orthorhombic system and the momentum $\vec{k}=(\pi,\pi,\pi)$ is invariant under $\mathcal{T}$. Therefore, one can simply use $\mathcal{T}$ at $\vec{k}=(\pi,\pi,\pi)$, instead of the combined symmetry $P\mathcal{T}$ in the main text, to form Kramer's pairing of the four intersecting band. Therefore, $S_{\text{\sg{19}}}^{\text{BI}}\subseteq 8\bN$.

\subsubsection{\sg{29} and \sg{33}}
Both \sg{29} ($Pca 2_1$) and \sg{33} ($Pna 2_1$) can be viewed as the combination of a $2_1$ screw along the z-axis and a glide about the $x$-$z$ plane. They contain the symmetry elements:
\begin{eqnarray}
S_z \equiv& T_{(0,0,1/2)} R_{z, \pi},\\
G_y \equiv& T_{(1/2,\tau_y, 0)} M_{y},\\
G_x \equiv& T_{(1/2,\tau_y, 1/2)} M_{x},
\end{eqnarray}
where the mirror $M_y = P R_{y, \pi}$ (recall $P$ denotes spatial inversion) is about $x$-$z$ plane, and $M_x$ is similarly defined. Note that $M_{\alpha}^2 = -1$ when acting on single-particle states. $\tau_y = 0$ and $1/2$ respectively for \sg{29} and \sg{33}. The following argument does not depend on the value of $\tau_y$, and therefore applies equally well to \sg{29} and \sg{33}.

Along the line $\vec k = (k_x, 0, \pi)$, $G_y$ is a symmetry and since $G_y^2 = T_{(1,0,0)}$, the eigenvalue of $G_y$ satisfies $\xi_y ^2 = - e^{- i k_x}$. At the TRIM $(0,0,\pi)$, $\xi_y= \pm i$ and the two bands in a Kramer's doublet have conjugate $\xi_y$. When these bands are evolved to $(\pi,0,\pi)$, they have $\xi_y = \pm 1$ and therefore the number of bands at $(\pi,0,\pi)$ with $\xi_y=1$ equals to that with $\xi_y = -1$.

A similar argument applies to the line $\vec k  = (\pi, 0, k_z)$. Here we observe $S_z^2 = T_{(0,0,1)}$ and $G_x^2= T_{(0,2 \tau_y, 1)}$, and therefore the respective eigenvalues satisfy $\xi_z^2 = \xi_x^2 = - e^{- i k_z}$. Combined with the argument in the preceding paragraph, we conclude that at $(\pi, 0, \pi)$ the number of bands with $\xi_{\alpha} = +1$ is the same as that with $\xi_{\alpha}=-1$ for $\alpha = x,y,z$.

In addition, the eigenvalues of a single band at $(\pi, 0,\pi)$ satisfy $\xi_z \xi_y \xi_x = 1$. This gives four 1D irreps: $(\xi_z, \xi_y, \xi_x) = (1,1,1)$, $(1,-1,-1)$, $(-1,1,-1)$ or $(-1,-1,1)$. 
Similar to the argument presented in the discussion for \sg{73} in the main text, the requirements of equal number of bands having $\xi_{\alpha} = \pm 1$ implies all four 1D irreps appear the same number of times in a set of isolated bands. 
Further noticing each of these 1D irreps is paired with another copy of itself under TR, we conclude $\mS^{\text{BI}}_{\mG} \subseteq 8\mathbb N $ for $\mG =$ \sg{29} and \sg{33}.

\section{V.~~~Band theoretical derivation of $\mS_{\mG}^{\text{BI}}$ for remaining space groups}
\label{V}
In the main text, we explained that the combinations of $\mS_{\mG}^{\text{AI}}$ and $\mS_{\Gamma}^{\text{BI}}$ ($\Gamma<\mG$) are sufficient to set $\mS_{\mG}^{\text{BI}}=\mS_{\mG}^{\text{AI}}$ for 218 out of 230 space groups. 

Among the 12 remaining space groups, let us first focus on the following eight space groups:
\begin{quote}
\sg{73}, \sg{106}, \sg{110}, \sg{133}, \sg{135}, \sg{142}, \sg{206}, and \sg{228}.
\end{quote}
It turns out they all satisfy $\mS_{\mG}^{\text{BI}}=\mS_{\mG}^{\text{AI}}=8\mathbb{N}$.  
The remaining 4 space groups are \sg{199}, \sg{214}, \sg{220} and \sg{230}, which we refer to as `Wyckoff-mismatched' and in general $\mS_{\mG}^{\text{BI}}\neq \mS_{\mG}^{\text{AI}}$ for them.

While the arguments presented can be simplified by making use of the tables in, say, Ref.~\cite{Bradley}, we aim to keep the discussion self-contained below. The only exception to this is the discussion for \sg{220}, in which we make use of the tabulated properties of the irreps in Refs.~\cite{Bradley,Bilbao}.

\subsection{A.~~~The eight remaining space groups with $\mS^{\text{BI}}_{\mG} = \mS^{\text{AI}}_{\mG} = 8 \mathbb N$: \\
\sg{73}, \sg{106}, \sg{110}, \sg{133}, \sg{135}, \sg{142}, \sg{206}, and \sg{228}}
To derive the tight constraints efficiently, we note the following: (i) all of the eight space groups have $\mS_{\mG}^{\text{AI}}=8\mathbb{N}$ \cite{ITC}; (ii) \sg{142}, \sg{206}, and \sg{228} contain \sg{73} as a $t$-subgroup and \sg{133} and \sg{135} contain \sg{106} as a $t$-subgroup. Here, $t$-subgroup just means that $v_{\mathcal{G},\mathcal{G'}}=1$. Hence, all we have to prove is $\mS_{\mG}^{\text{BI}}\subseteq 8\mathbb{N}$ for \sg{73}, \sg{106}, and \sg{110}, and we have already discussed \sg{73} in the main text. 

\subsubsection{\sg{106}}
\sg{106} $(P4_2bc)$ has 8 symmetry elements modulo the lattice translation $T$ and is generated by a $4_2$ screw $S_z$ and a glide $G_y$:
\begin{eqnarray}
S_z&\equiv& T_{(0,0,\frac{1}{2})}R_{z,\frac{\pi}{2}},\\
G_y&\equiv& T_{(\frac{1}{2},\frac{1}{2},0)}M_y,
\end{eqnarray}
where $M_y\equiv PR_{y,\pi}$ is the mirror reflection about $xz$ plane and $P$ is the inversion.  Unlike the previous example, \sg{106} belongs to the primitive tetragonal system and all momenta of the form $(k_x, k_y, k_z)$ with $k_x, k_y, k_z=0,\pm\pi$ are TRIMs.  

Along the line $\vec{k}=(\pi,\pi,k_z)$, $\mG_{\vec{k}}$ has all symmetry elements (i.e., $\mG_{\vec{k}}=\mG$) regardless of the value of $k_z\in[0, \pi]$.  There are two 2D irreps $U_{\vec{k}}^{(\xi)}$ ($\xi=\pm1$).  After an appropriate unitary transformation, we have
\begin{eqnarray}
U_{\vec{k}}^{(\xi)}(S_z)&=&\xi\frac{-i\sigma_0+\sigma_2}{\sqrt{2}}e^{-i\frac{k_z}{2}},\\
U_{\vec{k}}^{(\xi)}(G_y)&=&\sigma_1.
\end{eqnarray}

Now we consider the time-reversal symmetry $\mathcal{T}$. First, note that $U_{\vec{k}}^{(-)*}=U_{\vec{k}}^{(+)}$ at $\vec{k}=(\pi,\pi,0)$, meaninig that $U_{\vec{k}}^{(-)}$ is the time-reversal pair of $U_{\vec{k}}^{(+)}$. Namely, the representation is in total 4D (see Sec.~III~B):
\begin{eqnarray}
&&U_{\vec{k}}=\begin{pmatrix}
U_{\vec{k}}^{(+)}&0\\0&U_{\vec{k}}^{(-)}
\end{pmatrix},\,\,
U_{\vec{k},\mathcal{T}}=\textstyle\begin{pmatrix}0&-\sigma_0\\\sigma_0&0.\end{pmatrix}.
\end{eqnarray}

On the other hand, $U_{\vec{k}}^{(\xi)*}=U_{\vec{k}}^{(\xi)}$ at $\vec{k}=(\pi,\pi,0)$. Thus, in contrast to the situation at $\vec{k}=(\pi,\pi,0)$, the time-reversal pair of $U_{\vec{k}}^{(+)}$ ($U_{\vec{k}}^{(-)}$) is another $U_{\vec{k}}^{(+)}$ ($U_{\vec{k}}^{(-)}$) at $\vec{k}=(\pi,\pi,\pi)$, forming 4D irreps,
\begin{eqnarray}
&U_{\vec{k}}^{(+)}=\begin{pmatrix}
U_{\vec{k}}^{(+)}&0\\0&U_{\vec{k}}^{(+)}
\end{pmatrix},U_{\vec{k}}^{(-)}=\begin{pmatrix}
U_{\vec{k}}^{(-)}&0\\0&U_{\vec{k}}^{(-)}
\end{pmatrix},&\\
&U_{\vec{k},\mathcal{T}}=\textstyle\begin{pmatrix}0&-\sigma_0\\\sigma_0&0\end{pmatrix}.&
\end{eqnarray}
This time-reversal pair-switching explains why it is impossible to isolate $8n-4$ bands: in order to form the right time-reversal pairs at the two ends of the line $(\pi,\pi,k_z)$, an even number of $U_{\vec{k}}^{(+)}$ and the same number of $U_{\vec{k}}^{(-)}$ must participate in the band structure and intersect with each other along this line. This means that $8$ branches are the minimum building block of the band insulator, i.e., $\mS_{\text{\sg{106}}}^{\text{BI}}\subseteq8\mathbb{N}$.

\subsubsection{\sg{110}}
\sg{110} ($I4_1cd$) belongs to the body-centered tetragonal system and is generated by a $4_1$-screw and a glide:
\begin{eqnarray}
S_z&\equiv& T_{(0,\frac{1}{2},\frac{1}{4})}R_{z,\frac{\pi}{2}},\\
G_y&\equiv& T_{(0,0,\frac{1}{2})}M_y.
\end{eqnarray}

We focus on the line $\vec{k}=(\pi,\pi,k_z)$ again.  Along this line, $S_z^2$ and $G_yS_z$ generate the order-four group (modulo the lattice translation). They form four 1D reps:
\begin{eqnarray}
&&U_{\vec{k}}^{(\xi_1,\xi_2)}(S_z^2)=i\xi_1 e^{-i\frac{k_z}{2}},\notag\\
&&U_{\vec{k}}^{(\xi_1,\xi_2)}(G_yS_z)=i\xi_2 e^{-i\frac{3k_z}{4}},\label{app:110}
\end{eqnarray}
where $\xi_1,\xi_2=\pm1$.  Note that $U_{\vec{k}+(0,0,4\pi)}^{(\xi_1,\xi_2)}=D_{\vec{k}}^{(\xi_1,-\xi_2)}$; namely, when $k_z$ increases from $0$ to $4\pi$ along this line, the branch with $\xi_2=\pm 1$ becomes the one with $\xi_2=\mp1$.  However, $(\pi,\pi,4\pi)$ and $(\pi,\pi,0)$ differ only by a reciprocal lattice vector and are hence the same point in the Brillouin zone. 
[Remember, $(0,0,2\pi)$ is not a reciprocal lattice vector due to the body-centered structure].  Therefore, both $U_{\vec{k}}^{(\xi_1,\xi_2)}$ and $U_{\vec{k}}^{(\xi_1,-\xi_2)}$ must appear together and intersect with each other some point along the line $\vec{k}=(\pi,\pi,k_z)$.

At $\vec{k}=(\pi,\pi,0)$, $\mathcal{T}$ is a symmetry.  Since $U_{\vec{k}=(\pi,\pi,0)}^{(\xi_1,\xi_2)}$ and $U_{\vec{k}=(\pi,\pi,0)}^{(-\xi_1,-\xi_2)}$ are complex conjugate with each other, they form a time-reversal pair and must be degenerate at $\vec{k}=(\pi,\pi,0)$. Hence all of the four 1D reps $\xi_1,\xi_2=\pm1$ appear precisely the same number of times along this line.

At $\vec{k}=(\pi,\pi,\pi)$, $\mathcal{T}$ alone is not a symmetry, but the product $G_y\mathcal{T}$ is. Furthermore, $(G_y\mathcal{T})^2=\mathcal{T}^2M_y^2T_{(0,0,1)}=-1$ since $k_z=\pi$ and the antiunitary operator $G_y\mathcal{T}$ thus enforces a Kramers pairing.  To see this more explicitly, let us take the basis of the 1D representations at $\vec{k}=(\pi,\pi,\pi)$. It satisfies 
\begin{eqnarray}
S_z^2|\xi_1,\xi_2\rangle&=&\xi_1|\xi_1,\xi_2\rangle,\\
G_yS_z|\xi_1,\xi_2\rangle&=&\xi_2 e^{-i\frac{\pi}{4}}|\xi_1,\xi_2\rangle.
\end{eqnarray}
Writing $|\xi_1,\xi_2\rangle'\equiv G_y\mathcal{T}|\xi_1,\xi_2\rangle$, it can be readily shown that
\begin{eqnarray}
S_z^2|\xi_1,\xi_2\rangle'&=&\xi_1|\xi_1,\xi_2\rangle'\\
G_yS_z|\xi_1,\xi_2\rangle'&=&\xi_1\xi_2 e^{-i\frac{\pi}{4}}|\xi_1,\xi_2\rangle'.
\end{eqnarray}
Namely, $|\xi_1,\xi_2\rangle'$ has the same eigenvalues as $|\xi_1,\xi_1\xi_2\rangle$.  In particular, $|{+1,+1}\rangle$ and $G_y\mathcal{T}|{+1,+1}\rangle$ transform in the same representation but they cannot be the same state since $(G_y\mathcal{T})^2=-1$. Therefore, there must be an even number of $(\xi_1, \xi_2)=(+1,+1)$ states. Since all four 1D reps appear the same number of times, we have proven that $8$ bands are the minimum building blocks, i.e.~$\mS^{\text{BI}}_{\text{\sg{110}}} \subseteq 8 \mathbb N$.

\subsection{B.~~~The four Wyckoff-mismatched space groups: \\
\sg{199}, \sg{214}, \sg{220}, and \sg{230}}

According to Table~\ref{summary} of the main text, $\mS_{\mG}^{\text{AI}}=4\mathbb{N}\setminus\{4\}$ and $\mS_{\Gamma}^{\text{BI}}=4\mathbb{N}$ for \sg{199} and \sg{214}. Therefore, we have to determine if it is possible to realize a $\nu=4$ BI. 
In fact, the answer varies depending on whether the SU(2) symmetry is present or not.  In Ref.~\cite{us2}, we found $\nu=4$ BI in spin-orbit coupled insulators both for \sg{199} and \sg{214}. However, it is not possible to realize such an insulator without SOC as we discuss now. To show this, it's sufficient to discuss only \sg{199}, since \sg{214} contains \sg{199} as a $t$-subgroup.

\subsubsection{\sg{199}}
\sg{199} ($I2_13$) belongs to body-centered cubic system. It has (i) the three screws as in \sg{73} and (ii) three-fold rotation $C_3$ cyclically permuting $(x,y,z)$, which in total form an order $12$ group modulo the lattice translation. 
As we explained in Sec.~III~C, in the absence of SOC, we can think of the problem as spinless electrons. In that case, there are three 1D irreps and one 3D irrep at four points in the BZ $\vec{k}=(0,0,0), (2\pi,0,0), (0,2\pi,0), (0,0,2\pi)$.  All of the three 1D irreps realize the screw trivially. Namely,
\begin{equation}
U_{\vec{k}}^{\text{1D}}(S_i)=+1\quad\text{for}\quad i=x,y,z.
\end{equation}
On the other hand, the 3D irrep reads
\begin{eqnarray}
U_{\vec{k}}^{\text{3D}}(S_x)=\text{diag}(+1,-1,-1),\\
U_{\vec{k}}^{\text{3D}}(S_y)=\text{diag}(-1,+1,-1),\\
U_{\vec{k}}^{\text{3D}}(S_z)=\text{diag}(-1,-1,+1).
\end{eqnarray}
Now let us consider the line from $(0,0,0)$ to $(0,0,2\pi)$. Along this line $S_z$ remains symmetry and its eigenvalue continuously changes by the factor of $e^{-i k_z/2}$. 
In particular, if one uses a 1D rep with $S_z=+1$ at $\vec{k}=(0,0,0)$, there must be at least one 3D rep at $\vec{k}=(0,0,2\pi)$ to account for the $S_z=-1$ eigenvalue. This argument proves that at least three bands, each doubly degenerate due to spin, must appear together and cross with each other for the spinless case. Therefore, for spinful electrons without SOC, we have proved $\nu\geq 6$, excluding the $\nu=4$ BI.

\subsubsection{\sg{220}}
For \sg{220}, we know that $\mS_{\text{\sg{220}}}^{\text{BI}}\supseteq\mS_{\text{\sg{220}}}^{\text{AI}}=4\mathbb{N}\setminus\{4,8,20\}$ and also that $\mS_{\text{\sg{220}}}^{\text{BI}}\subseteq\mS_{\text{\sg{24}}}^{\text{BI}}=4\mathbb{N}$ as \sg{220} contains \sg{24} as a $t$-subgroup. Therefore, all we have to check is if $\mS_{\text{\sg{220}}}^{\text{BI}}$ contains 4, 8, and/or 20 or not.  The answer again depends on whether the spin SU(2) symmetry present or not. 

\paragraph{Without SU(2) spin-rotation symmetry}
As demonstrated in Ref.~\cite{us2}, there are filling-enforced quantum band insulators at filling $\nu = 8$ and $20$. We now prove that it is impossible to realize a BI at $\nu=4$.

We start by studying the irreps at $\vec k = (2 \pi, 0, 0)\equiv \text{H}$. 
There are two 2D irreps ($\Gamma^{\text{H}}_6, \Gamma^{\text{H}}_7$) and one 4D irrep ($\Gamma^{\text{H}}_8$) (See Table~\ref{tab:220SpinfulPH}, which we reproduced based on Ref.~\cite{Bradley}). Under TR, the two 2D irreps are paired to form a 4D co-representation $\Gamma^{\text{H}}_6 \oplus \Gamma^{\text{H}}_7$, and two copies of the 4D irrep are paired to form a 8D co-representation $2\Gamma^{\text{H}}_8$~\cite{2015arXiv151200074W,Bradley}. For $\nu=4$, therefore, the filled bands must correspond to $\Gamma^{\text{H}}_6 \oplus \Gamma^{\text{H}}_7$. 

Along the line P-H ($\vec k = (1-\kappa) \text{P} + \kappa \text{H}$, where $\text{P} = (\pi,\pi,\pi)$), the little group (modulo lattice translation) is given by $\mG_{\vec k}=\{(1), (7), (10), (14), (17), (23)\}$, with $(j)$ denoting the $j$-th element as listed in Ref.~\cite{ITC}. In Table~\ref{tab:220SpinfulPH} we list the relevant symmetry characters for the irreps involved. 
(Note that all the listed characters are subjected to a $\pm 1$ ambiguity arising from the $-1$ phase picked up by an electron under a $2\pi$ rotation. This ambiguity, however, is `global' in the sense that one simply picks a convention for each of the elements in the space group, and the compatibility relations we discuss are independent of such choice of convention.)
Despite each row represents a different irrep, two rows for a high symmetry point can look identical due to the restriction to a subgroup of the little group. Observe that the proposed $\nu=4$ band insulator must correspond to $2 \Gamma^{\text{PH}}_6$ along the line P-H, which after sending $\kappa: 1\rightarrow 0$ corresponds to $2 \Gamma^{\text{P}}_6$  at P.

\begin{center}
\begin{table}
\caption{Relevant symmetry characters for the irreps along the line P-H. We let $\theta \equiv e^{ i \pi/4}$. We use the same labeling as Ref.~\cite{Bradley}.
\label{tab:220SpinfulPH}
}
\begin{tabular}{c|c|cccccc}
$\vec k$ & Irrep & (1) & (7) & (10) & (14) & (17) & (23)\\
\hline
H & $\Gamma_6^{\text{H}}$ & $2$ & $1$  & $-1$ & $0$ & $0$ & $0$ \\
$(\kappa=1)$ & $\Gamma_7^{\text{H}}$  & $2$ & $1$  & $-1$ & $0$ & $0$ & $0$\\
~& $\Gamma_8^{\text{H}}$  & $4$ & $-1$  & $1$ & $0$ & $0$ & $0$ \\
\hline
PH & $\Gamma_4^{\text{PH}}$  & $1$ & $-1$  & $-\theta^{4 \kappa}$ & $\theta^{ 5 \kappa -1}$ & $ \theta^{ \kappa-1}$ & $ \theta^{ \kappa-1 }$\\
$(1-\kappa)\text{P} + \kappa \text{H}$& $\Gamma_5^{\text{PH}}$ & $1$ & $-1$  & $-\theta^{4 \kappa}$ & $- \theta^{5 \kappa- 1}$ & $-\theta^{\kappa-1}$ & $- \theta^{ \kappa-1}$\\
~& $\Gamma_6^{\text{PH}}$ & $2$ & $1$  & $\theta^{4 \kappa}$ & $0$ & $0$ & $0$\\
\hline
P & $\Gamma_4^{\text{P}}$  & $1$ & $-1$  & $-1$ & $ \theta^*$ & $ \theta^*$ & $\theta^* $\\
$(\kappa=0)$& $\Gamma_5^{\text{P}}$ & $1$ & $-1$  & $-1$ & $-\theta^* $ & $-\theta^* $ & $-\theta^*$\\
~& $\Gamma_6^{\text{P}}$& $2$ & $1$  & $1$ & $0$ & $0$ & $0$\\
~& $\Gamma_7^{\text{P}}$ & $3$ & $0$ & $0$ & $\theta^*$ & $\theta^*$ & $\theta^*$\\
~& $\Gamma_8^{\text{P}}$ & $3$ & $0$ & $0$ & $-\theta^*$ & $-\theta^*$ & $-\theta^*$\\
\hline
\end{tabular}
\end{table}
\end{center}

Next we perform a similar analysis along the line N-P ($\vec k = (1-\kappa) \text{N} + \kappa \text{P}$ with $\text{N} = (\pi,\pi,0)$), for which we have $\mG_{\vec k} = \{(1),(2),(13),(14)\}$ (Table~\ref{tab:220SpinfulNP}). In particular, observe that the symmetry character of (2) picks up a phase of $\theta^{2} = i$ going from P to N $(\kappa:1\rightarrow 0)$, and therefore the character of (2) at N is $-4 i$ for the proposed $\nu=4$ band insulator. Since N is a TRIM, this is contradictory to the assumed TR invariance, implying $4 \not \in \mS^{\text{BI}}_{\text{\sg{220} }}$

\begin{center}
\begin{table}
\caption{Relevant symmetry characters for the irreps along the line N-P. We let $\theta \equiv e^{i \pi/4}$. We use the same labeling as Ref.~\cite{Bradley}.
\label{tab:220SpinfulNP}
}
\begin{tabular}{c|c|cccc}
$\vec k$ & Irrep & (1) & (2) & (13) & (14) \\
\hline
P & $\Gamma_4^{\text{P}}$ & $1$ & $-1$  & $-\theta^*$ & $\theta^*$ \\
$(\kappa=1)$& $\Gamma_5^{\text{P}}$ & $1$ & $-1$  & $\theta^*$ & $-\theta^*$ \\
~& $\Gamma_6^{\text{P}}$ & $2$ & $-2$  & $0$ & $0$ \\
~& $\Gamma_7^{\text{P}}$ & $3$ & $1$  & $-\theta^*$ & $\theta^*$ \\
~& $\Gamma_8^{\text{P}}$ & $3$ & $1$  & $\theta^*$ & $-\theta^*$ \\
\hline
NP
& $\Gamma_1^{\text{NP}}$ & $1$ & $   \theta^{- 2 (1+\kappa)}$  & $ \theta^{- \kappa}$ & $ \theta^{-  (2+3 \kappa)} $\\
$(1-\kappa)\text{N} + \kappa \text{P}$  & $\Gamma_2^{\text{NP}}$& $1$ & $  \theta^{2(1-\kappa)  }$  & $\theta^{- \kappa}$ & $ \theta^{2-  3  \kappa } $\\
~& $\Gamma_3^{\text{NP}}$& $1$ & $ \theta^{- 2 (1+\kappa)  }$  & $- \theta^{-  \kappa  }$ & $  \theta^{2 -3  \kappa  } $\\
~& $\Gamma_4^{\text{NP}}$& $1$ & $ \theta^{2 (1-\kappa)  }$  & $-\theta^{-  \kappa}$ & $  \theta^{- (2+3 \kappa)} $\\
\hline
N & $\Gamma_1^{\text{N}}$ & $1$ & $-i $  & $ 1$ & $ -i $\\
$(\kappa=0)$  & $\Gamma_2^{\text{N}}$ & $1$ & $i$  & $ 1$ & $i  $\\
~ & $\Gamma_3^{\text{N}}$& $1$ & $-i $  & $-1$ & $ i $\\
~ & $\Gamma_4^{\text{N}}$ & $1$ & $i $  & $ -1$ & $ -i  $\\
\hline
\end{tabular}
\end{table}
\end{center}

\paragraph{In the presence of SU(2) symmetry}
BIs at filling either $\nu=4$, $8$, or $20$ are not allowed in the absence of SOC, i.e., $\mS_{\text{\sg{220}}}^{\text{BI}}=4\mathbb{N}\setminus\{4,8,20\}$ as we show now.
This follows from what we call `the compatibility condition', which we have already used many times so far.  In general, suppose that $g\in\mG$ is a symmetry along a high-symmetry line $\vec{k}_\kappa=(1-\kappa)\vec{k}_1+\kappa\vec{k}_2$ connecting two high-symmetry momenta $\vec{k}_1$ and $\vec{k}_2$.
We denote the eigenvalues of $g$ by $\eta_i(\kappa)$ ($i=1,2,\ldots$) at each $\kappa$. Then, the number of occurrence of the eigenvalue $\eta_i(\kappa)$ must be a constant along this line because of the continuity of the band structure.  This condition restricts the allowed combinations of irreps at $\vec{k}_1$ and $\vec{k}_2$. 

We list in Table~\ref{app_tab_220} the number and the dimension of irreps of $\mathcal{G}_{\vec{k}}$ at each high-symmetry momentum for spinless electrons with $\mG=\text{\sg{220}}$. For example, there are in total five irreps at $\Gamma$ (two 1Ds, one 2D, and two 3Ds). 
One can find the full list of these irreps in Ref.~\cite{Bilbao} and we follow the labeling there.  After imposing the compatibility conditions and the time-reversal symmetry, we found that there are essentially three building blocks of band insulators. They are composed of $6$, $8$, and $12$ branches, respectively labeled by $[1]$, $[2]$ and $[3]$ below. Here we show the number of occurrence of irreps at high-symmetry momenta, for each of the three building blocks:
\begin{eqnarray}
{[1]}\quad\Gamma&:& (n_1,1-n_1,1,n_1,1-n_1)\quad (n_1=0,1)\notag\\
\text{H}&:& (0,0,0,1,1)\notag\\
\text{P}&:& (n_2,1-n_2,1)\quad (n_2=0,1)\notag\\
\text{PA}&:& (n_3,1-n_3,1)\quad (n_3=0,1)\notag\\
\text{D}&:& (3).\\
{[2]}\quad\Gamma&:& (n_1,2-n_1,0,n_1,2-n_1)\quad (n_1=0,1,2)\notag\\
\text{H}&:& (1,1,0,1,1)\notag\\
\text{P}&:& (0,0,2)\notag\\
\text{PA}&:& (0,0,2)\notag\\
\text{D}&:& (4).\\
{[3]}\quad\Gamma&:& (0,0,0,2,2)\notag\\
\text{H}&:& (1,1,2,1,1)\notag\\
\text{P}&:& (n_1,2-n_1,2)\quad (n_1=0,1,2)\notag\\
\text{PA}&:& (n_2,2-n_2,2)\quad (n_2=0,1,2)\notag\\
\text{D}&:& (6).
\end{eqnarray}

These blocks can together form band insulators with filling $\nu\in\{6n_1+8n_2+12n_3|n_i\in\mathbb{N}\}=2\mathbb{N}\setminus\{2,4,10\}$ for spinless electrons symmetric under $\mG=\text{\sg{220}}$ and the time-reversal $\mathcal{T}$.  For spinful electrons with SU(2) spin rotation symmetry, each band is doubly degenerate and therefore $\mS_{\text{\sg{220}}}^{\text{BI}}=4\mathbb{N}\setminus\{4,8,20\}$.

\subsubsection{\sg{230}}
Since \sg{230} contains \sg{73} as a $t$-subgroup, we know $\mS_{\text{\sg{230}}}^{\text{BI}}\subseteq\mS_{\text{\sg{73}}}^{\text{BI}}=8\mathbb{N}$. Combining this with $\mS_{\text{\sg{230}}}^{\text{BI}}\supseteq\mS_{\text{\sg{230}}}^{\text{AI}}=8\mathbb{N}\setminus\{8\}$, the remaining question is if it is possible to isolate $8$ bands.  Again, we constructed a tight-biding model that realizes a $\nu=8$ BI in Ref.~\cite{us2} for spin-orbit coupled case. However, similarly to \sg{199} and \sg{214}, it is impossible to achieve a BI at $\nu=8$ without SOC.  This follows from the fact that $8\not \in \mS^{\text{BI}}_{\text{\sg{220}}\times\text{SU}(2)}$ as 
\sg{220} is a $t$-subgroups of \sg{230}.

\begin{table}
\begin{center}
\caption{\label{app_tab_220}The irreps of $\mathcal{G}_{\vec{k}}$ for spinless electrons with $\mG=\text{\sg{220}}$.}
\begin{tabular}{c|c|l}
$\vec{k}$ & Number of irreps & Dimension of each irrep \\\hline
$\Gamma$: $(0,0,0)$ & $5$  & $(1,1,2,3,3)$\\
H: $(2\pi,0,0)$ & $5$  & $(1,1,2,3,3)$\\
P: $(\pi,\pi,\pi)$ & $3$  & $(2,2,4)$\\
PA: $(\pi,\pi,-\pi)$ & $3$  & $(2,2,4)$\\
D: $(\pi,\pi,0)$ & $1$  & $(2)$\\
\end{tabular}
\end{center}
\end{table}

\section{VI.~~~Band Insulators on Compact Flat Manifolds}
\label{SMVI}
In the main text, we discussed how to put a system of spinful electrons on a compact flat manifold through the example of \sg{73}. Here we expand it with a more general discussion.

\subsection{A.~~~Putting system on compact flat manifolds}
Consider a system of electrons symmetric under $\mG$ and choose a Bieberbach subgroup $\Gamma \subseteq \mG$.  
The corresponding manifold $\mathcal{M}=\mathbb{R}^3/\Gamma$ corresponds to one of the 10 compact flat manifolds in 3D.  There are two equivalent approaches to put the system originally defined in $\mathbb{R}^3$ onto $\mathcal{M}$.

(1) In the first approach, electronic creation operators transform projectively under $g\in\mG$:
\begin{equation}
\hat{g}\hat{c}_i^\dagger(\vec{r})\hat{g}^{-1}=\hat{c}_j^\dagger(g(\vec{r}))(U_g^{(0)})_{ji}.
\end{equation}
Here $U_g^{(0)}$ is a $\mathbb{Z}_2$-projective representation of $\mG$ that satisfies $U_g^{(0)}U_g'^{(0)}=\omega_{g,g'}^{(0)}U_{gg'}^{(0)}$ with $\omega_{g,g'}^{(0)}$ being $z(p_g,p_{g'})\in\mathbb{Z}_2$ in Eq.~\eqref{app:pr_spin}.
We assume the Hamiltonian and commutation relations are all invariant under this symmetry operation.   In addition, we also assume the fermion parity symmetry $\hat{c}_i^\dagger(\vec{r})\rightarrow (-1)^{\hat{F}}\hat{c}_i^\dagger(\vec{r})(-1)^{\hat{F}}=-\hat{c}_i^\dagger(\vec{r})$, which is equivalent to say each $U_g$ has a sign ambiguity.  If one uses $U_g=\xi_gU_g^{(0)}$ ($\xi_g=\pm1$) instead, the factor system becomes 
\begin{equation}
\omega_{g,g'}=\frac{\xi_g \xi_{g'}}{\xi_{gg'}}\omega_{g,g'}^{(0)}.
\label{app:omega2}
\end{equation}

For a fixed-point-free subgroup $\Gamma$ of $\mG$, we can choose $\xi_g$ in such a way that $\omega_{\gamma,\gamma'}=1$ for all $\gamma,\gamma'\in\Gamma$. Namely, $U_g$ is a linear representation of $\Gamma$. Once we have such $U_g$, one can put the system on the compact flat manifold $\mathcal{M}$ by identifying positions and operators respectively by
\begin{eqnarray}
\vec{r}&\sim&\gamma(\vec{r}),\\
\hat{c}^{\dagger}_i(\vec{r})&\sim&\hat{\gamma} \hat{c}^{\dagger}_i(\vec{r}) \hat{\gamma}^{-1} = \hat{c}^{\dagger}_j(\gamma(\vec{r})) (U_\gamma)_{ji}.
\end{eqnarray}

(2) An alternative approach is to use the double group (see Sec.~VII~B).  In this view, electrons follow a \emph{linear} representation of $\mG^F$, the `doubled' version of $\mG$. 
An element of $\mG^F$ is $g_\xi=(\xi,g)$, where $\xi\in\mathbb{Z}_2$ and $g\in\mG$. For example, the `$2\pi$-rotation', which was identity $e\in\mG$ in the above projective representation, now corresponds to $(-1,e)$ in this language. The product of $g_\xi, g'_{\xi'}\in\mG^F$ is defined as $(\xi,g) (\xi',g') = (\omega_{g,g'}^{(0)}\xi\xi',gg')$. Note that $\mG^F$ is in general a nontrivial $\mathbb Z_2$ extension of $\mG$, and $\omega_{g,g'}^{(0)}$ encodes the group structure. For instance, one has $(+1, R_{x,\pi}) (+1, R_{x,\pi})  = (-1, e) \neq ((+1)^2, R^2_{x,\pi}) $. 

We assume electrons transform linearly under $\mG^F$:
\begin{equation}
\hat{g}_\xi\hat{c}_i^\dagger(\vec{r})\hat{g}_\xi^{-1}=\hat{c}_j^\dagger(g(\vec{r}))(\mathcal{U}_{g_\xi})_{ji},
\end{equation}
where $\mathcal{U}_{g_\xi}$ is a linear representation of $\mG^F$ that satisfies $\mathcal{U}_{e_\xi}=\xi$ and $\mathcal{U}_{g_{+1}}=U_g^{(0)}$.  We assume the Hamiltonian and commutation relations are all unchanged under $\mG^{F}$.  There is a natural two-to-one projection $\pi: \mG^F \rightarrow \mG: g_\xi \mapsto g$ but there does not exist a unique map from $\mG\rightarrow \mG^F$.  However, in order to consistently mod-out, we need to specify a homomorphism $\epsilon: \Gamma \rightarrow \Gamma^F$ satisfying $\pi\circ \epsilon = \text{Id}_{\Gamma}$. Here, $\Gamma^F$ is the doubled version of $\Gamma$, which is just the corresponding restriction of $\mG^F$, i.e., $(\xi,\gamma)$ ($\xi\in\mathbb{Z}_2$ and $\gamma\in\Gamma$). Once we have such $\epsilon$, one can put the system on the compact flat manifold $\mathcal{M}$ by identifying positions and operators by
\begin{eqnarray}
\vec{r}&\sim&\gamma(\vec{r}),\\
\hat{c}^{\dagger}_i(\vec{r})&\sim&\hat{\epsilon}(\gamma) \hat{c}^{\dagger}_i(\vec{r}) \hat{\epsilon}(\gamma)^{-1} = \hat{c}^{\dagger}_j(\gamma(\vec{r})) (\mathcal{U}_{\epsilon(\gamma)})_{ji},
\end{eqnarray}
and $\mathcal{U}_{\epsilon(\gamma)}=\xi_{{\epsilon(\gamma})}U_\gamma^{(0)}$ by definition.

Let us explain a little more about the meaning of the homomorphism $\epsilon$. Writing $\epsilon(\gamma)=(\xi_{\epsilon(\gamma)},\gamma)$, we have
\begin{eqnarray}
\epsilon(\gamma)\epsilon(\gamma')&=&(\omega_{\gamma,\gamma'}^{(0)}\xi_{\epsilon(\gamma)}\xi_{\epsilon(\gamma')},\gamma\gamma'),\\
\epsilon(\gamma\gamma')&=&(\xi_{\epsilon(\gamma\gamma')},\gamma \gamma').
\end{eqnarray}
Therefore, for $\epsilon$ to be a homomorphism, we need
\begin{equation}
\frac{\xi_{\epsilon(\gamma)}\xi_{\epsilon(\gamma')}}{\xi_{\epsilon(\gamma\gamma')}}\omega_{\gamma,\gamma'}^{(0)}=1.
\end{equation}
Comparing with Eq.~\eqref{app:omega2}, one can see that the requirement of homomorphism is nothing but $\omega_{\gamma,\gamma'}=1$ for all $\gamma,\gamma'\in\Gamma$ in the previous approach.

\subsection{B.~~~Remnant symmetries}
An element $g$ in $\mG$ but not in $\Gamma$ may or may not remain a symmetry on $\mathcal{M}$. The general necessary and sufficient condition is that
\begin{eqnarray}
\forall\gamma\in\Gamma,\,\, g\gamma g^{-1}\in\Gamma\,\,\text{and}\,\, U_gU_\gamma U_g^{-1} = U_{g\gamma g^{-1}}.\label{remantsymm}
\end{eqnarray}
For the time reversal symmetry $\mathcal{T}$, one should check 
\begin{eqnarray}
U_{\mathcal{T}}U_\gamma^* U_{\mathcal{T}}^{-1} = U_{\gamma},
\end{eqnarray}
where $U_{\mathcal{T}}$ is unitary and satisfies $U_{\mathcal{T}} U_{\mathcal{T}}^{*} =-1$. As long as one deals with $\mathbb{Z}_2$-extension, this is always true as $\mG$ and $\mathcal{T}$ commute. 

Equivalently, we require the symmetry $g$ to be compatible with the chosen homomorphism $\epsilon$ involved in modding-out, i.e.~we require $g \epsilon(\gamma) g^{-1} \in \epsilon(\Gamma)$ for all $\gamma \in \Gamma$. In addition, since the spatial action of $\hat \epsilon(\gamma)$ is trivialized after modding-out, $\hat \epsilon(\gamma)$ effectively becomes an on-site unitary operator. 

\subsection{C.~~~$S_{\mG}^{\text{BI}}$ for fixed-point-free space groups $\Gamma$}
Here we present an alternative derivation of $S_{\Gamma}^{\text{BI}}$ by putting band insulators with $\mG=\Gamma$ on the corresponding Bieberbach manifold.  We will show that $S_{\Gamma}^{\text{BI}}=2|\Gamma/T_{\Gamma}|\mathbb{N}$ for a fixed-point-free space group $\Gamma$, where $T_{\Gamma}$ is the translation subgroup of $\Gamma$ and $|\Gamma/T_{\Gamma}|$ is the number of elements of the quotient group $\Gamma/T_{\Gamma}$.

When $\Gamma$ is just the translation subgroup $\text{\sg{1}}\sim\mathbb{Z}^3$, the compact manifold $\mathcal{M}=\mathbb{R}^3/\Gamma=T^3$ (the three torus) always contains an integer number of unit cells. However, this is not the case for the other 9 Bieberbach space groups - the number of unit cells on the manifold $\mathcal{M}$ is a integer multiple of the fraction
\begin{equation}
\frac{1}{|\Gamma/T_{\Gamma}|}.
\end{equation}
To see this, let us construct $\mathcal{M}$ in two steps. First, get the three torus $T^3=\mathbb{R}^3/T_{\Gamma}$ using the translation subgroup $T_{\Gamma}$ of $\Gamma$. Suppose that $T^3$ contains $L^3$ unit cells.  Then, further identify points of $T^3=\mathbb{R}^3/T_{\Gamma}$ using the remnant symmetries $\Gamma'\equiv\Gamma/T$ on $T^3$. One then gets the manifold 
\begin{equation}
T^3/\Gamma'=(\mathbb{R}^3/T_\Gamma)/(\Gamma/T_\Gamma)=\mathbb{R}^3/\Gamma=\mathcal{M}.
\end{equation}
Clearly, the number of unit cells on $\mathcal{M}=T^3/\Gamma'$ is reduced from that of $T^3$ by the factor of $|\Gamma'|=|\Gamma/T_{\Gamma}|$. Thus the number of unit cells on $\mathcal{M}$ is $\frac{L^3}{|\Gamma/T_{\Gamma}|}$. 
We list $|\Gamma/T_{\Gamma}|$ for each $\Gamma$ in Table~\ref{tab_app_uc}. 

Since $\nu$ is the number of electrons per unit cell, the total number of electrons on $\mathcal{M}$ is given by $N_{\mathcal{M}}=\nu \frac{L^3}{|\Gamma/T_{\Gamma}|}$.  
If $N_{\mathcal {M}}$ is odd, the single particle spectrum on $\mathcal{M}$ has at least two-fold degeneracy due to Kramer's paring. Thus $N_{\mathcal{M}}$ must be an even integer in order to be a unique gapped ground state.  Since we can choose $L^3$ to be an odd integer that is co-prime with $|\Gamma/T_{\Gamma}|$, $\nu$ must be an integer multiple of $2|\Gamma/T_{\Gamma}|$.  By comparing this result with $\mathcal{S}_{\Gamma}^{\text{AI}}$ in Table~\ref{tab_app_bi2}, one gets $\mathcal{S}_{\Gamma}^{\text{BI}}= 2|\Gamma/T_{\Gamma}|\mathbb{N}$ for Bieberbach space groups.

As an example, let us discuss $\Gamma=\text{\sg{19}}$, relevant for $\mathcal{G}=\text{\sg{73}}$ discussed in the main text.
$\Gamma=\text{\sg{19}}$ is generated by three screws $\tilde{S}_\alpha = T_{\vec \tau_{\alpha}}^L R_{\alpha,\pi}$ with $\alpha = x,y,z$. $R_{\alpha,\theta}$ represents the anti-clockwise rotation by angle $\theta$ around the positive $\alpha$-axis and $\vec \tau_{x} = (1/2,1/2,0)$, $\vec \tau_{y} = (0,1/2,1/2)$ and $\vec \tau_{z} = (1/2,0,1/2)$.  The translation subgroup $T_\Gamma$ is generated by 
\begin{eqnarray}
\tilde{S}_x^2=T_{\hat{x}}^L&=&(L,0,0),\\
\tilde{S}_y^2=T_{\hat{y}}^L&=&(0,L,0),\\
\tilde{S}_z^2=T_{\hat{z}}^L&=&(0,0,L).
\end{eqnarray}
Hence, the three torus $T^3=\mathbb{R}^3/T_{\Gamma}$ contains $L^3$ unit cells.  The coset $\Gamma/T_\Gamma$ is the order four group $\{[e], [\tilde{S}_x], [\tilde{S}_y], [\tilde{S}_z]\}$ with $[\tilde{S}_x]^2=[T_{\hat{x}}^L]=[e]$ and $[\tilde{S}_x][\tilde{S}_y]=[T_{\hat{z}}^{-L}\tilde{S}_z]=[\tilde{S}_z]$. Therefore, the manifold $\mathcal{M}=\mathbb{R}^3/\Gamma$ contains $\frac{L^3}{|\Gamma/T_\Gamma|}=\frac{L^3}{4}$ unit cells.  Here, a unit cell of \sg{19} is spanned by $T_{\hat{\alpha}}$ and has the unit volume.

Now, recall that \sg{73} includes the body-centered translation $T_{(\frac{1}{2},\frac{1}{2},\frac{1}{2})}$.  The volume of the unit cell of \sg{73} is thus $\frac{1}{2}$.  In other words, a single unit cell of \sg{19} contains two unit cells of \sg{73}. Hence, $\mathcal{M}=\mathbb{R}^3/\Gamma$ contains $\frac{2L^3}{4}$ unit cells of \sg{73}.  

More generally, if $\Gamma\leq\mathcal{G}$, the manifold $\mathcal{M}=\mathbb{R}^3/\Gamma$ contains 
\begin{equation}
\frac{L^3}{|\Gamma/T_\Gamma|}v_{\mathcal{G},\Gamma} =L^3\frac{|T_{\mathcal{G}}/T_{\Gamma}|}{|\Gamma/T_\Gamma|}
\end{equation}
unit cells of $\mathcal{G}$. Hence, the interacting bound derived in Ref.~\cite{us1} for spinful electrons can be expressed as
\begin{eqnarray}
S_{\mathcal{G}}&\equiv&2m_{\mathcal{G}}\mathbb{N},\\
m_{\mathcal{G}}&\equiv&\max_{\Gamma\leq\mathcal{G}}\frac{|\Gamma/T_\Gamma|}{|T_{\mathcal{G}}/T_{\Gamma}|}=\max_{\Gamma\leq\mathcal{G}}\frac{|\mathcal{G}/T_{\mathcal{G}}|}{|\mathcal{G}/\Gamma|}.
\end{eqnarray}
$S_{\mathcal{G}}$ coincides with $\cap_{\Gamma \leq \mG} (\mS_{\Gamma}^{\text{BI}}/v_{\mG,\Gamma})$ in Table~\ref{summary} of the main text.

\section{VII.~~~Projective Representation}
\label{SMVII}
Here we briefly review the basics of projective representations, which appeared many times in this paper.

\subsection{A.~~~Definitions}
When a set of matrices $U_{\mathrm{g}}$ satisfy $U_{\mathrm{g}}U_{\mathrm{g}'}=\omega_{\mathrm{g},\mathrm{g}'} U_{\mathrm{g}\mathrm{g}'}$ for all $\mathrm{g},\mathrm{g}'\in G$, we call $U_{\mathrm{g}}$ a projective representation of a group $G$ and $\omega_{g,g'}\in A\subseteq\text{U(1)}$ the factor system.  To respect the associative property of matrix product, the factor system must satisfy the cocycle condition $\omega_{\mathrm{g},\mathrm{g}'}\omega_{\mathrm{g}\mathrm{g}',\mathrm{g}''}=\omega_{\mathrm{g},\mathrm{g}'\mathrm{g}''}\omega_{\mathrm{g}',\mathrm{g}''}$.

The factor system $\omega$ intrinsically has ambiguity originating from the redefinition freedom $U_{\mathrm{g}}\rightarrow U_{\mathrm{g}}a_{\mathrm{g}}$ ($a_{\mathrm{g}}\in A$). $\omega$ and $\omega'$ are thus said to be equivalent when there exists a map $a: G\rightarrow A$ such that $\omega_{\mathrm{g},\mathrm{g}'}'=\omega_{\mathrm{g},\mathrm{g}'}\frac{a_{\mathrm{g}}a_{\mathrm{g}'}}{a_{\mathrm{g}\mathrm{g}'}}$. Inequivalent factor systems are fully classified by $H^2(G,A)$.  One can set $\omega_{\mathrm{e},\mathrm{g}}=\omega_{\mathrm{g},\mathrm{e}}=1$ ($\mathrm{e}$ is the identity of $G$) without loss of generality.

\subsection{B.~~~Double group}
Projective representations are sometimes treated by linear representations of an enlarged group (the `double' group when $A=\mathbb{Z}_2$). To explicitly see the relation, consider a group $\tilde{G}=A\times_\omega G$ that is a product $A\times G$ as a set (so $|\tilde{G}|=|A||G|$) but is endowed with the multiplication:
\begin{equation}
(a,\mathrm{g})(a',\mathrm{g}')=(\omega_{\mathrm{g},\mathrm{g}'}aa',\mathrm{g}\mathrm{g}'),\,\,\, a,a'\in A,\,\mathrm{g},\mathrm{g}'\in G.
\end{equation}
The short exact sequence $1\rightarrow A\xrightarrow{\iota} \tilde{G}\xrightarrow{\pi} G\rightarrow1$, where $\iota$ is the injection $a\mapsto (a,\mathrm{e})$ and $\pi$ is the projection $(a,\mathrm{g})\mapsto \mathrm{g}$, defines a central extension of $G$ by $A$.  Let $\mathcal{U}$ be a linear irreducible representation of $\tilde{G}$ [i.e., $\mathcal{U}_{(a,\mathrm{g})}\mathcal{U}_{(a,\mathrm{g})}=\mathcal{U}_{(a,\mathrm{g})(a',\mathrm{g}')}$] that satisfies $\mathcal{U}_{(a,\mathrm{e})}=a\openone$.  Then a projective irreducible representation of $G$ is given by $U_{\mathrm{g}}\equiv \mathcal{U}_{(1,\mathrm{g})}$. This is how one can go back and forth between $\mathcal{U}$ and $U$.

\subsection{C.~~~Properties of projective irreps}
Let us restrict ourself to the case of finite $G$ and write irreducible representations as $U^{(\alpha)}$ ($\alpha=1,2,\ldots,N$; $N$ is the number of distinct irreps).  Then the following useful relations hold~\cite{GK1,GK2}:
\begin{eqnarray}
&&\sum_{\alpha=1}^{N}\text{dim}[U^{(\alpha)}]^2=|G|,\\
&&\sum_{g\in G}\text{tr}[U^{(\alpha)}_{\mathrm{g}}]^*\,\text{tr}[U^{(\beta)}_{\mathrm{g}}]=|G|\delta_{\alpha,\beta},\\
&&N=\frac{1}{|G|}\sum_{\mathrm{g},\mathrm{g}'\in  G}\frac{\omega_{\mathrm{g},\mathrm{g}'}}{\omega_{\mathrm{g}',\mathrm{g}}}\delta_{\mathrm{g}\mathrm{g}',\mathrm{g}'\mathrm{g}}.
\end{eqnarray}
Note that 1D representations do not always exist for projective representations unlike linear (i.e., non projective) representations for which the trivial representation $U_{\mathrm{g}}=1$ is always valid.
1D representations are allowed only when $\omega_{\mathrm{g},\mathrm{g}'}=\omega_{\mathrm{g}',\mathrm{g}}$ for all $\mathrm{g}, \mathrm{g}'$.

\end{document}